\documentclass[iop]{emulateapj}

\usepackage{graphicx}
\usepackage{amssymb}
\usepackage{epstopdf}
\usepackage{natbib}
\usepackage{subfigure}
\usepackage{multirow}
\usepackage{verbatim}
\usepackage[normalem]{ulem}
\shorttitle{INFRARED SEDS FROM HERSCHEL}
\shortauthors{LEE ET AL.}

\begin{document}

\title{Multi-wavelength SEDs of {\em Herschel} Selected Galaxies in the COSMOS Field}
\author{Nicholas Lee\altaffilmark{1}, D. B. Sanders\altaffilmark{1}, Caitlin M. Casey\altaffilmark{1}, N.Z. Scoville\altaffilmark{2}, Chao-Ling Hung \altaffilmark{1}, Emeric Le Floc'h\altaffilmark{3}, Olivier Ilbert\altaffilmark{4}, Herv\'{e} Aussel\altaffilmark{3}, Peter Capak \altaffilmark{2,5}, Jeyhan S. Kartaltepe\altaffilmark{6}, Isaac Roseboom\altaffilmark{7}, Mara Salvato\altaffilmark{8,9}, M. Aravena\altaffilmark{10,11}, J. Bock\altaffilmark{2,12}, S. J. Oliver\altaffilmark{7}, L. Riguccini\altaffilmark{13,14}, M. Symeonidis\altaffilmark{15,16}}
\altaffiltext{1}{Institute for Astronomy, 2680 Woodlawn Dr., Honolulu, HI, 96822, USA}
\altaffiltext{2}{California institute of Technology, MS 105-24, 1200 East California Boulevard, Pasadena, CA 91125}
\altaffiltext{3}{UMR AIM (CEA-UP7-CNRS), CEA-Saclay, Orme des Merisiers, b\^{a}t. 709, F-91191 Gif-sur-Yvette Cedex, France}
\altaffiltext{4}{Aix Marseille Universit\'e, CNRS, LAM (Laboratoire d'Astrophysique de Marseille) UMR 7326, 13388, Marseille}
\altaffiltext{5}{Spitzer Science Center, 314-6 Caltech, 1201 East California Boulevard, Pasadena, CA 91125}
\altaffiltext{6}{National Optical Astronomy Observatory, 950 North Cherry Ave., Tucson, AZ, 85719, USA}
\altaffiltext{7}{Institute for Astronomy, University of Edinburgh, Royal Observatory, Blackford Hill, Edinburgh EH9 3HJ}
\altaffiltext{8}{Max-Planck for Extraterrestrial Physics, Giessenbachstrasse 1, 85748 Garching, Germany}
\altaffiltext{9}{Cluster of Excellence, Boltzmann Strasse 2, 85748, Garching, Germany}
\altaffiltext{10}{European Southern Observatory, Alonso de C\'{o}rdova 3107, Vitacura Santiago, Chile}
\altaffiltext{11}{Universidad Diego Portales, Faculty of Engineering, Av. Ej\'{e}rcito 441, Santiago, Chile}
\altaffiltext{12}{Jet Propulsion Laboratory, 4800 Oak Grove Drive, Pasadena, CA 91109, USA}
\altaffiltext{13}{NASA Ames Research Center, Moffett Field, CA, USA}
\altaffiltext{14}{BAER Institute, Santa Rosa, CA, USA}
\altaffiltext{15}{University of Sussex, Department of Physics and Astronomy, Pevensey 2 Building, Falmer, Brighton BN1 9QH, Sussex, UK}
\altaffiltext{16}{Mullard Space Science Laboratory, University College London, Holmbury St. Mary, Dorking, Surrey RH5 6NT, UK}

\begin{abstract}

We combine {\em Herschel} PACS and SPIRE maps of the full 2 deg$^{2}$ COSMOS field with existing multi-wavelength data to obtain template and model-independent optical-to-far-infrared spectral energy distributions (SEDs) for 4,218 {\em Herschel}-selected sources with log($L_{\rm{IR}}/L_{\odot}$) = 9.4--13.6 and $z = 0.02$--3.54.  Median SEDs are created by binning the optical to far-infrared (FIR) bands available in COSMOS as a function of infrared luminosity.  {\em Herschel} probes rest-frame wavelengths where the bulk of the infrared radiation is emitted, allowing us to more accurately determine fundamental dust properties of our sample of infrared luminous galaxies.  We find that the SED peak wavelength ($\lambda_{\rm{peak}}$) decreases and the dust mass ($M_{\rm{dust}}$) increases with increasing total infrared luminosity ($L_{\rm{IR}}$).  In the lowest infrared luminosity galaxies (log($L_{\rm{IR}}/L_{\odot}) = 10.0$--11.5), we see evidence of Polycyclic Aromatic Hydrocarbons (PAH) features ($\lambda \sim 7$--9 $\mu$m), while in the highest infrared luminosity galaxies ($L_{\rm{IR}} > 10^{12} L_{\odot}$) we see an increasing contribution of hot dust and/or power-law emission, consistent with the presence of heating from an active galactic nucleus (AGN).  We study the relationship between stellar mass and star formation rate of our sample of infrared luminous galaxies and find no evidence that {\em Herschel}-selected galaxies follow the $SFR/M_{*}$ ``main sequence'' as previously determined from studies of optically selected, star-forming galaxies.  Finally, we compare the mid-infrared (MIR) to FIR properties of our infrared luminous galaxies using the previously defined diagnostic, IR8~$\equiv L_{\rm{IR}} / L_{8}$, and find that galaxies with $L_{\rm{IR}}  \gtrsim 10^{11.3} L_{\odot}$ tend to systematically lie above ($\times 3$--5) the IR8 ``infrared main sequence'', suggesting either suppressed PAH emission or an increasing contribution from AGN heating.
\end{abstract}

\keywords{galaxies: evolution - galaxies: high-redshift - infrared: galaxies}

\section{Introduction}

Surveys of the sky in far-infrared wavelengths are crucial to gain a complete understanding of the nature of extragalactic objects.  Dust in galaxies absorbs the ultra-violet (UV) radiation from young, massive stars and reradiates thermally in the infrared.  This happens to some degree in all galaxies, but in the most bolometrically luminous galaxies, the total infrared radiation ($L_{\rm{IR}} \equiv L(8$--1000 $\mu$m))  dominates the total emission \citep[e.g.][]{1996ARA&A..34..749S}.  Although rare locally, these Luminous Infrared Galaxies (LIRGs, $10^{11} L_{\odot} < L_{\rm{IR}} < 10^{12} L_{\odot}$) and Ultra-Luminous Infrared Galaxies (ULIRGs, $L_{\rm{IR}} > 10^{12} L_{\odot}$) have a significant contribution to the buildup of stellar mass and growth of supermassive black holes at higher redshifts, producing as much as 50\% of the stellar mass in the universe at redshifts $z \sim 2$--3 \citep{2005ApJ...622..772C,2005ApJ...632..169L,2012ApJ...761..140C} and as much as $\sim 30$\% of the integrated black hole growth through highly obscured accretion \citep{2009ApJ...706..535T,2010ApJ...722L.238T}.  

The {\em Herschel Space Observatory} \citep{2010A&A...518L...1P} provides us with the first sensitive observations of the sky at far-infrared wavelengths, where the bulk of the infrared radiation is emitted.  This allows us to more accurately measure bolometric infrared luminosity, which is a fundamental property of galaxies and is thought to be an excellent tracer of the star formation rate in (U)LIRGs \citep[e.g.][]{1998ARA&A..36..189K}.  One of the key deep fields surveyed with {\em Herschel} was the \emph{Cosmic Evolution Survey} (COSMOS) field \citep{2007ApJS..172....1S}, a 2 deg$^{2}$ field with extensive coverage of multiwavelength imaging and spectroscopic observations. With this new and unique dataset, we can begin to examine what powers the extreme luminosities in infrared luminous galaxies. 


A major drawback of many previous studies of infrared luminous galaxies is that they do not have sufficient wavelength coverage at FIR wavelengths ($\sim$100~$\mu$m), where the SED peaks.  Many previous studies have relied on uncertain extrapolations from available wavebands (e.g. {\em Spitzer} $24 \mu$m) that can be affected by AGN contamination or poor assumptions of the SED shape in the FIR.  This becomes especially problematic for studies at high redshift, as the available wavebands are redshifted even farther away from the FIR peak of the SED, and the MIR bands begin to be contaminated by PAH emission that can vary substantially in strength.  In addition, many of the extrapolations are based on fits to SED libraries \citep[e.g.][]{Chary:2001p1425,2002ApJ...576..159D,2007ApJ...657..810D} that are constructed from galaxies at low redshift, and these models may not represent the SEDs of high redshift galaxies.  Indeed, studies using stacking techniques to better study long wavelength data have shown that extrapolations from mid-IR wavelengths are generally accurate at low redshifts and low infrared luminosities, but become significantly less accurate at high redshifts and high infrared luminosities \citep{Papovich:2007p38,2010ApJ...717..175L}.  Studies at submillimeter wavelengths ($\lambda \gtrsim 850 \mu$m) avoid the problem of extrapolation from MIR wavelengths, but are affected by a severe bias in dust temperature \citep{2004ApJ...611...52B,2004ApJ...614..671C,2009MNRAS.399..121C}.  

In this paper, we select galaxies at wavelengths where their FIR SEDs peak, avoiding the need for uncertain extrapolations from MIR observations and avoiding the temperature bias of submillimeter surveys.  We study the full SEDs (UV-to-FIR) of a large population of {\em Herschel}-selected galaxies without using any prior assumptions of SED shape.  

This paper is organized as follows: the data are described in \S \ref{sec:data} and in \S\ref{sec:meas_lir} we examine the fundamental dust properties of our full sample of 4,218 galaxies.  In \S\ref{sec:med_sed}, we construct median SEDs and investigate how they evolve as a function of $L_{\rm{IR}}$.  In \S\ref{sec:discussion}, we discuss whether there is evidence that our objects lie on the optical and infrared ``main-sequence'' as suggested by previous studies of star-forming galaxies and infrared luminous galaxies.  When calculating rest-frame quantities, we use a cosmology with $\Omega_{m} = 0.3$, $\Lambda = 0.7$, and $H_{0} = 70$ km s$^{-1}$ Mpc$^{-1}$ and we also assume a Salpeter initial mass function \citep[IMF:][]{1955ApJ...121..161S} when deriving SFRs and stellar masses.

\section{Data and Sample Selection}\label{sec:data}

\subsection{Far-Infrared Observations}

We use observations from the {\em ESA Herschel Space Observatory} \citep{2010A&A...518L...1P}, in particular employing {\em Herschel's} large telescope and powerful science payload to do photometry using the Photodetector Array Camera and Spectrometer \citep[PACS,][]{2010A&A...518L...2P} and Spectral and Photometric Imaging Receiver \citep[SPIRE,][]{2010A&A...518L...3G} instruments.  

The COSMOS field has been observed down to the confusion limit of $\sim 20$mJy at $250 \mu$m, $350 \mu$m, and $500 \mu$m by {\em Herschel} SPIRE as part of the {\em Herschel} Multi-tiered Extragalactic Survey \citep[HerMES;][]{2012MNRAS.424.1614O}.  In order to measure accurate flux densities of sources in the confusion-dominated SPIRE mosaics, we use the linear inversion technique of cross-identification (hereafter XID), as described in \citet[][]{2010MNRAS.409...48R,2012MNRAS.419.2758R}.  A linear inversion technique fits the flux density of all known sources simultaneously, but the accuracy of such a technique is greatly dependent on the completeness of the sample of prior known sources. For infrared sources, the list of priors is generally taken as sources that are bright at $24 \mu$m or 1.4 GHz.  Observations at both of these wavelengths have much better resolution than FIR observations, and both are thought to correlate strongly with total infrared luminosity \citep{1985ApJ...298L...7H,1992ARA&A..30..575C,1998ARA&A..36..189K,2009ApJ...692..556R}.  Our COSMOS-{\em Spitzer} \citep{2007ApJS..172...86S} survey provides very deep coverage at $24 \mu$m, containing $>$39,000 sources with $S_{24} > 80 \mu$Jy (90\% completeness limit) over an effective area of 1.68 deg$^{2}$ \citep{2009ApJ...703..222L}, and complete coverage at 1.4 GHz with the VLA, leading to 2,865 sources with $1 \sigma \sim 12 \mu$Jy \citep{2010ApJS..188..384S}.  While we include 1.4 GHz counterparts when extracting sources from the SPIRE map to ensure completeness of our list of priors, we also require a $24 \mu$m counterpart for our SPIRE detections in order to maintain uniformity with the PACS sources.  The XID method provides flux density measurements for every input source, but following the recommendations in \citet{2010MNRAS.409...48R}, we only keep sources with $S/N > 5$ and $\chi ^{2} < 5$.  This yields 8,308 sources detected at $250 \mu$m, 3,186 sources detected at $350 \mu$m, and 955 sources detected at $500 \mu$m, with typical $1 \sigma$ total noise (instrumental $+$ confusion) of 2.2, 2.9, and 3.2 mJy in the 250, 350, and $500 \mu$m bands, similar to what was found in \citet{2012MNRAS.419.2758R}.

{\em Herschel} PACS observations in the COSMOS field at $100 \mu$m and $160 \mu$m were performed as a part of the PACS Evolutionary Probe program \cite[PEP;][]{2011A&A...532A..90L}.  Catalog extraction was performed blindly, using the Starfinder PSF-fitting code \citep{2000A&AS..147..335D}, and employing $24 \mu$m priors following the method described in \citet{2009A&A...496...57M}, but in this work we use the sources extracted from $24 \mu$m priors to maintain consistency with the SPIRE sources that are extracted using $24 \mu$m priors.  

\subsection{Observations at Shorter Wavelengths}\label{sec:opt_match}

Optical counterparts of our {\em Herschel} sample are found following the matching algorithm adopted by \citet{2009ApJ...703..222L}.  As a necessity of our {\em Herschel} source extraction, all of our sources have $24 \mu$m counterparts.  Thus, we first find $K_{s}$-band counterparts \citep[from][]{McCracken:2009p2464} for each $24 \mu$m source using a matching radius of $2''$.  A matching radius of $1''$ is then used to match these $K_{s}$-band counterparts to the $i^{+}$-band selected photo-z catalog of \cite{2009ApJ...690.1236I}.  Sources without $K_{s}$-band counterparts are matched directly to the photo-z catalog using a matching radius of 2''.  The photo-z catalog contains {\em Subaru B, V, g+, r+, i+, z+,} \citep{2007ApJS..172...99C,2007ApJS..172....9T} and Ultra-VISTA {\em J, K} \citep{2012A&A...544A.156M} fluxes, in addition to excellent photometric redshifts for 933,789 sources selected at $i^{+}_{AB} < 26.5$ mag from the $Subaru$/Suprime-CAM observations of COSMOS \citep{2007ApJS..172...99C}.  {\em Spitzer} IRAC \citep{2007ApJS..172...86S} $3.6 \mu$m, $4.5 \mu$m, $5.8 \mu$m, and $8 \mu$m counterparts are found using a similar method, by matching to the $K_{s}$-band counterparts using a $1''$ matching radius.  We also find sub-millimeter flux densities from AzTEC \citep{2008MNRAS.385.2225S,2011MNRAS.415.3831A} and MAMBO \citep{2007ApJS..172..132B} by using a matching radius of $2''$ for the sources that have radio counterparts.  In all, we have full multi-wavelength coverage of {\em Herschel}-selected sources from optical through sub-millimeter wavelengths.  

\subsection{Photometric Redshifts}

As mentioned in \S\ref{sec:opt_match}, the extensive multiwavelength coverage in the COSMOS field leads to extremely accurate photometric redshifts (hereafter photo-z), detailed in \citet{2009ApJ...690.1236I}.  Photo-z are calculated using fluxes in 30 bands, covering the far-UV at 1550 \AA \ to the mid-IR at $8.0 \mu$m.  The uncertainties in the photo-z depend primarily on the redshift and apparent $i^{+}$ magnitude of the source, with errors increasing with fainter and more distant galaxies, but a comparison with faint spectroscopic samples in the COSMOS field revealed a dispersion as low as $\sigma_{\Delta z / (1+z_{s})} = 0.06$ for sources with 23 mag $< i^{+}_{\mathrm{AB}} < 25$ mag at $1.5 \lesssim z \lesssim 3$ \citep{2007ApJS..172...70L}.  \citet{2012ApJ...761..140C} find that photo-z's can be much less accurate for infrared-selected samples, with $\sigma_{\Delta z / (1+z_{s})} \approx 0.3$ at $z < 2$, but there is no evidence of a systematic offset.  

Approximately 1000 of our $24 \mu$m sources are also detected in the X-ray by XMM-\emph{Newton} \citep{2010ApJ...716..348B} or {\em Chandra} \citep{2011ApJ...741...91C}, and for these sources we use the photo-z's derived from \citet{2011ApJ...742...61S}, who have the best photometric redshifts ever produced for AGN, reaching $\sigma_{\Delta z / (1+z_{s})} = 0.015$ for sources with $i^{+}_{\mathrm{AB}} < 22.5$ out to $z = 4.5$.    

\subsection{{\em Herschel} Sample Selection} \label{sec:selection}

We restrict our sample of {\em Herschel}-selected galaxies to sources with $\ge 5 \sigma$ detections in at least two of the five {\em Herschel} PACS and SPIRE bands (100, 160, 250, 350, or $500 \mu$m).  This requirement reduces the contamination from spurious detections and gives us enough data points in the far-infrared to accurately constrain the shape of the infrared SED (see \S\ref{sec:meas_lir}).  This selection ultimately results in 4,218 sources spanning redshifts $0.02 < z < 3.5$. 

\begin{figure}[!h]
\centering
\includegraphics[width=8 cm]{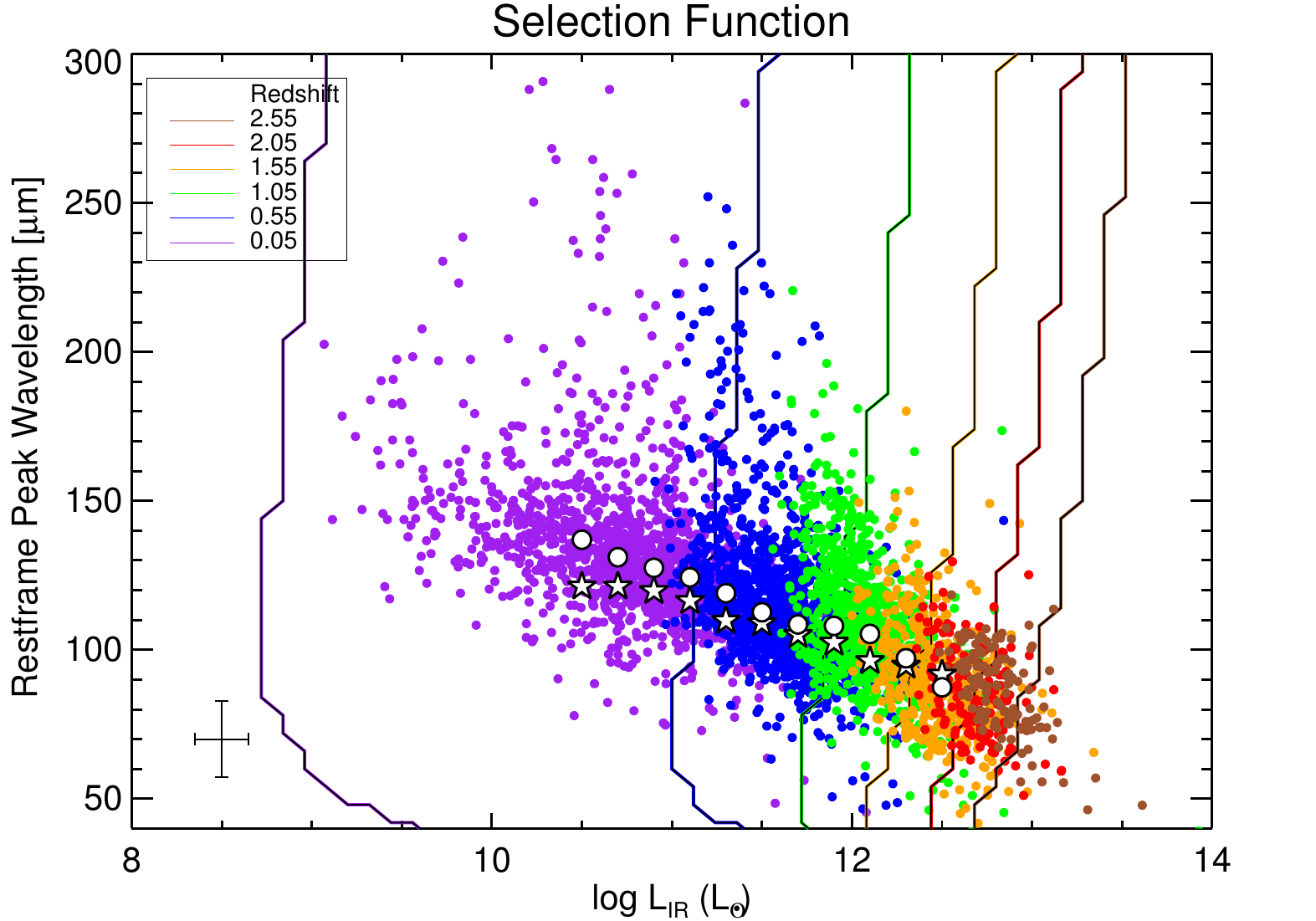}
\caption{Selection function from our criteria requiring at least two detections in the {\em Herschel} PACS \& SPIRE bands.  The lines represent the selection function at various redshifts.  The measured peak wavelengths and infrared luminosities of our {\em Herschel} sample are over-plotted as colored dots, with each color corresponding to galaxies that are at similar redshifts to the lines.  The white circles represent the median peak wavelengths of our sample, while the white stars represent the mean peak wavelengths from the \citet{2013MNRAS.431.2317S} study of {\em Herschel} galaxies.  Even though our selection function is slightly biased against cold dust temperature sources, we find that our sample of {\em Herschel} galaxies have slightly longer median peak wavelengths than the \citet{2013MNRAS.431.2317S} study.}
\label{fig:selection_func}
\end{figure}

As with any selection, there are biases that affect our sample.  The requirement of a $24 \mu$m source to use as a prior biases our sample against heavily obscured objects and high-redshift objects \citep{2010MNRAS.409...48R,2012ApJ...761..140C} or galaxies with strong silicate absorption features that are redshifted into the $24 \mu$m band \citep{2011A&A...534A..15M}.  The biases introduced by our requirement of $5 \sigma$ detections in two of the five {\em Herschel} bands is less obvious.  To explore the biases introduced by our selection criteria, we model the SEDs of galaxies spanning a wide range of peak wavelength ($\lambda_{\rm{peak}} \sim 10$--$300 \mu$m), infrared luminosity (log($L_{\rm{IR}}/L_{\odot}) \sim 8$--14), and redshift ($z \sim 0$--3.5).  We then convolve these (redshifted) model SEDs with the transmission curves of the relevant {\em Spitzer} and {\em Herschel} bands to determine their observed flux densities.  Figure \ref{fig:selection_func} displays the selection functions in peak wavelength and infrared luminosity space at different redshifts.  At most redshifts, we see a slight bias against sources with long peak wavelengths (colder dust temperatures), although by requiring detections in multiple wavelengths, we see much flatter selection functions than typically seen in single-band selections.

\section{Dust Properties} \label{sec:meas_lir}

We estimate the total infrared luminosity ($L_{\rm{IR}}$) of individual {\em Herschel}-selected galaxies by directly fitting their FIR photometry to a coupled modified greybody plus a MIR power law\footnote{http://www.ifa.hawaii.edu/~cmcasey/sedfitting.html}, as in C12.  The main strength of this technique is that we do not rely on templates which incorporate a myriad of free parameters, most of which cannot realistically be constrained by data; instead, using this simple model, we can cleanly fit the available photometric data without introducing biases which are template-dependent. 

C12 compare this greybody fitting technique with template fits from \citet[][hereafter CE01]{Chary:2001p1425}, \citet{2002ApJ...576..159D}, \citet{2007ApJ...657..810D} and \citet{Siebenmorgen:2007p1876} using local (U)LIRGs from the GOALS survey \citep[][U et al. 2012]{2009PASP..121..559A}, and find that the simple greybody plus power-law fits provide a statistically better fit to the data than any of the templates at {\em all} wavelengths.  One potential drawback of this simple fit is that it does not fit PAH features in the MIR, although such spectral features could be modeled on top of the simple fit in the future.  However, the net contribution of PAH features to the integrated 8--$1000 \mu$m infrared luminosity is negligible ($< 5$\%).  Without detailed information in the MIR, fits to PAH features can be extremely uncertain.  

As suggested in C12, we fix the MIR powerlaw slope, $\alpha = 2.0$, and the dust emissivity, $\beta = 1.5$.  For each source we then perform the fit using all available photometric data at observed wavelengths $\lambda \ge 24 \mu$m.  Our SED fits allow us to constrain $L_{\rm{IR}}$, peak wavelength ($\lambda_{\rm{peak}}$), and dust mass ($M_{\rm{dust}}$), with typical uncertainties of $\sigma_{L_{\rm{IR}}} \sim 0.15$ dex, $\sigma_{\lambda_{\rm{peak}}} \sim 13 \mu$m, and $\sigma_{M_{\rm{dust}}} \sim 0.38$ dex.  The rest-frame SED peak wavelength of $S_{\nu}$ is a proxy for dust temperature ($\lambda_{\rm{peak}} \propto 1/T_{\rm{dust}}$), but the actual conversion to dust temperature is very dependent on the assumed opacity and emissivity model.  For example, an SED that peaks at $100 \mu$m can have a dust temperature of 29$^\circ$K (blackbody), 31$^\circ$K (optically thin greybody), 44$^\circ$K (greybody with $\tau = 1$ at $100 \mu$m), or 46$^\circ$K (greybody with $\tau = 1$ at $200 \mu$m).  Due to these uncertainties in dust temperature from different model assumptions (see C12 for more details), we prefer to estimate $\lambda_{\rm{peak}}$, which is insensitive to model assumptions.   

We plot $\lambda_{\rm{peak}}$ and $L_{\rm{IR}}$ for our entire sample in Figure \ref{fig:selection_func} on top of the selection function for our sample.  \citet{2013MNRAS.431.2317S} previously analyzed the dust temperatures of a sample of {\em Herschel}-selected galaxies that is near-complete in SED types and is thought to be representative of the infrared galaxy population as a whole up to $z \sim 2$.  We convert their measured mean dust temperatures to $\lambda_{\rm{peak}}$ and find good agreement between our two studies: both studies show similar trends in rest-frame peak wavelength vs. $L_{\rm{IR}}$, with nearly equal values at the highest infrared luminosities and slightly cooler values for our current study at lower infrared luminosities.

\begin{figure}[!h]
\centering
\includegraphics[width=8 cm]{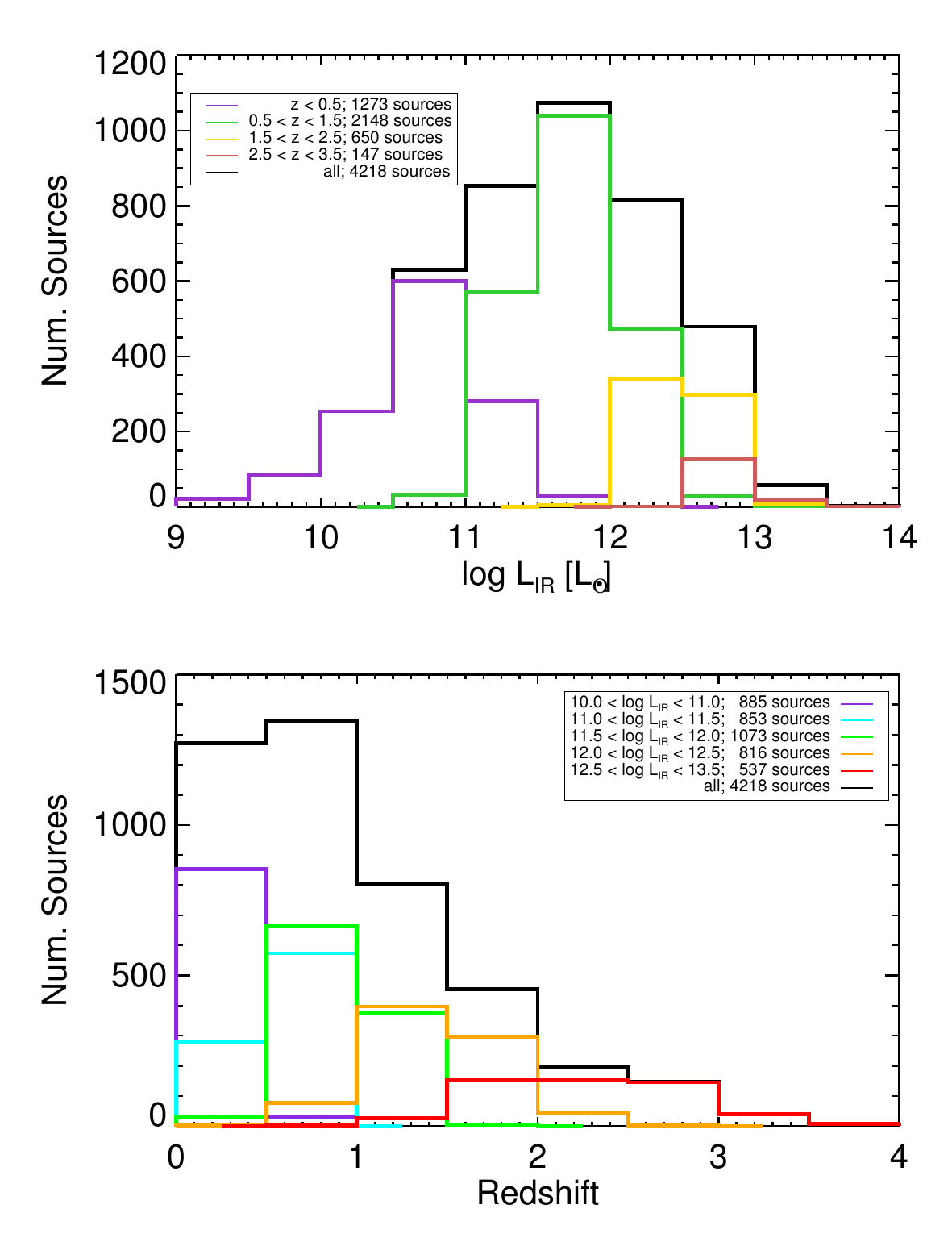}
\caption{Histograms of both bolometric infrared luminosity (top) and redshift (bottom) of our sample of over 4000 {\em Herschel} galaxies selected in the COSMOS field, requiring at least two detections in the far-infrared {\em Herschel} PACS or SPIRE bands.  We also plot the histograms of sources split into redshift bins and infrared luminosity bins.  The luminosity bins are selected so that there are approximately equal numbers of sources in each bin.}
\label{fig:lir_hist}
\end{figure}

\begin{figure}[!h]
\centering
\includegraphics[width=8 cm]{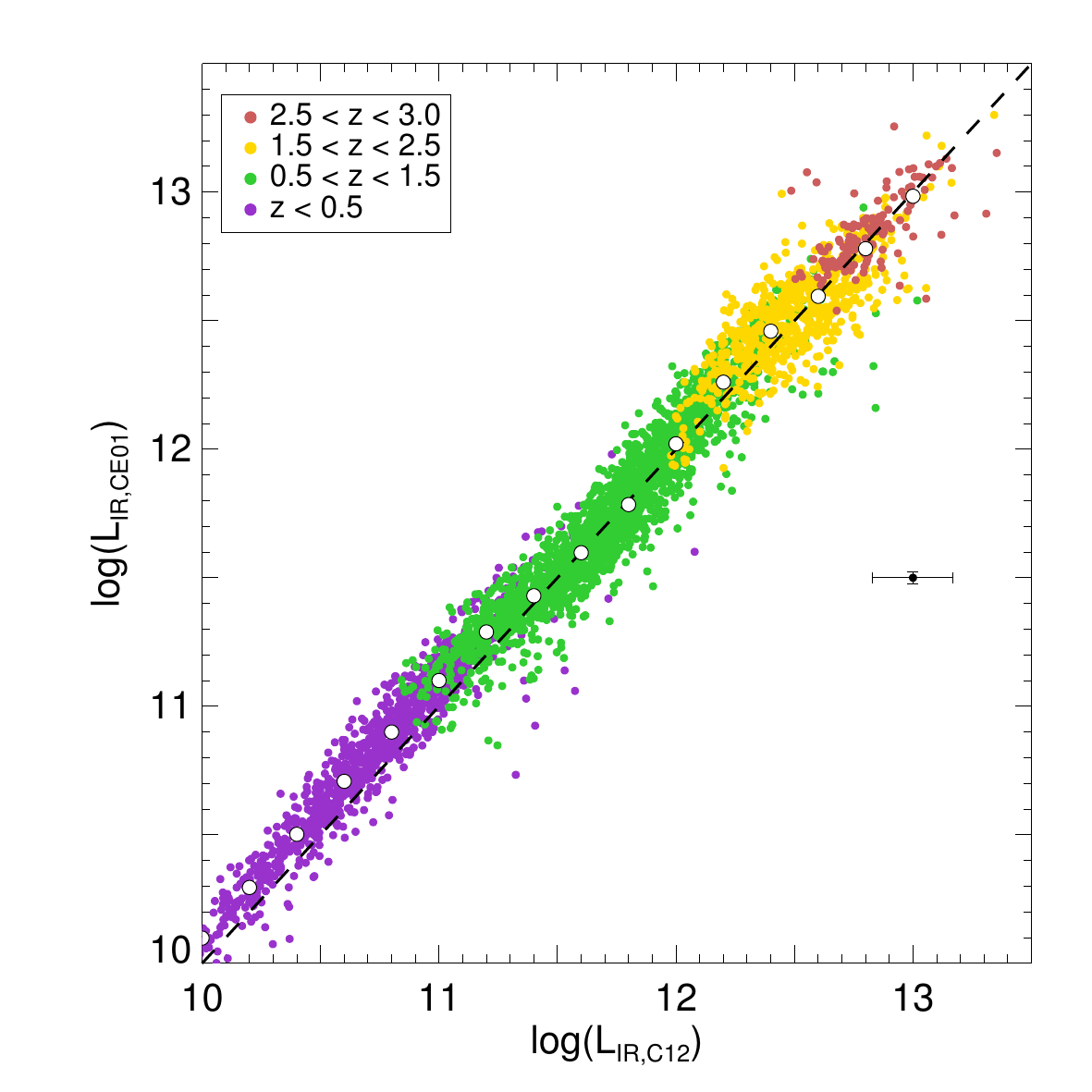}
\caption{A comparison of the total infrared luminosity ($L_{\rm{IR}}$) derived from our greybody fits on the x-axis and the $L_{\rm{IR}}$ derived from fits to template libraries from \citet{Chary:2001p1425} on the y-axis.  Different colored filled circles represent galaxies in different redshift ranges, spanning $0 < z < 3.5$, while white circles represent running median values of the distribution.  Typical errors are displayed on the right side of the plot.  The $L_{\rm{IR}}$ values measured from the two methods agree well, with the largest offset at low redshifts/infrared luminosities, where the C12 greybody fits have systematically lower infrared luminosities (although still within the typical uncertainties).}
\label{fig:lir_casey_vs_ce}
\end{figure}

Our sample of {\em Herschel}-selected galaxies spans a range log$(L_{\rm{IR}}/L_{\odot}) \approx 9.4$--13.6, which we show in Figure \ref{fig:lir_hist}.  We then compare $L_{\rm{IR}}$ determined from our greybody \& power-law fits to $L_{\rm{IR}}$ determined from fitting to SED templates from CE01 in Figure \ref{fig:lir_casey_vs_ce}.  We use the code \emph{Le Phare}\footnote{http://www.cfht.hawaii.edu/~arnouts/lephare.html} developed by S. Arnouts and O. Ilbert to fit the infrared SEDs to the models from CE01.  \emph{Le Phare} is a data analysis package used primarily to compute photometric redshifts, but can also be used to perform a $\chi^{2}$ analysis to find best fit galactic model SEDs in the far-infrared.  At low redshift (and low infrared luminosity), we see that the C12 greybody fits measure lower infrared luminosities than the CE01 template fits by an average of about 0.1 dex, similar to what was seen in C12 when they compared these two methods using the GOALS sample of local (U)LIRGs.  Over the entire sample, the luminosities computed from each method differ by an average of $\sim 0.11$ dex, within the typical uncertainty in $L_{\rm{IR}}$ of $\sim 0.15$ dex, with a slightly larger disagreement at high redshifts and high infrared luminosities.  This is not surprising since all of our sources are detected in multiple {\em Herschel} bands and are fairly well constrained at FIR wavelengths.  

\section{Median SEDs}\label{sec:med_sed}

To examine detailed galaxy SEDs without making assumptions on the SED shape, we construct average SEDs that will allow us to study the average properties of a population of galaxies.  By combining galaxies at different redshifts, we also sample different rest-frame wavelengths, allowing us to sample a larger portion of the full SED without requiring observations in additional passbands.  In this section, we describe our methodology for constructing median SEDs and then discuss what these SEDs tell us about the average properties of infrared luminous galaxies.
 
\subsection{Constructing Median SEDs}

An inherent assumption in any averaging technique is that all of the sources in a particular bin have similar SEDs.  We restrict our averaging analysis to galaxies with similar emission properties by splitting our sample of {\em Herschel}-selected galaxies into bins based on their total infrared luminosity calculated from their individual SED fits.  The implications of this binning are discussed in \S\ref{sec:selection_effects}.  We split our sample into five infrared luminosity bins to probe SED evolution with SFR: log$(L_{\rm{IR}}/L_{\odot}) = 10$--10.99, 11--11.49, 11.5--11.99, 12--12.49, and 12.5-13.5.   Due to the dynamical range of this survey, these bins also coincide with different redshift bins, with the lowest luminosity sources at $\langle z \rangle \sim 0.3$ and the highest at $\langle z \rangle \sim 2$.  

\begin{figure}[!h]
\centering
\includegraphics[width=8 cm]{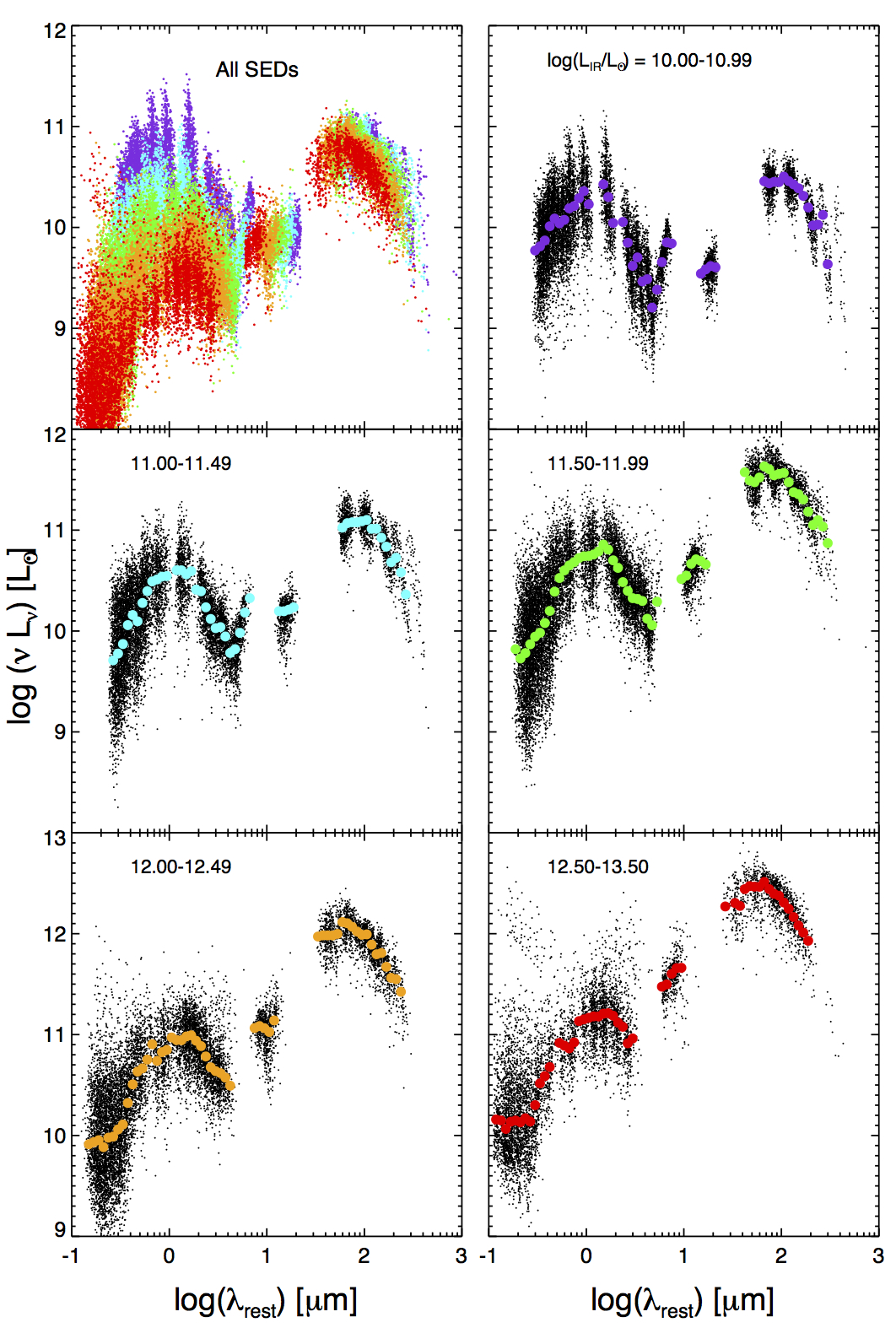}
\caption{Spectral Energy Distributions (SEDs) of all sources in our sample.  The top left panel displays the SEDs of the full sample of galaxies, in units of $\nu L_{\nu}$, redshifted to rest frame wavelengths and normalized so that every galaxy has $L_{\rm{IR}} = 10^{11} L_{\odot}$, colored by luminosity bin.  The remaining plots display the SEDs of the galaxies in each infrared luminosity bin, normalized to the mean $L_{\rm{IR}}$ in that bin (black points).  Overplotted on each plot is the derived median SED that was calculated in each bin (colored points).  Error bars for the median SEDs can be found in Figure \ref{fig:med_sed}.}
\label{fig:all_sed}
\end{figure}

For each source, we combine all the available photometry in COSMOS, from {\em Subaru B}-band through {\em Herschel}-SPIRE $500 \mu$m for all sources (or up to $\approx 1$ mm for those sources detected with AzTEC and/or MAMBO).  The SEDs for individual sources are redshifted to the object rest-frame wavelengths and converted to units of $\nu L_{\nu}$ using photometric redshifts.  In each luminosity bin, we then normalize all the SEDs so that each source has the same infrared luminosity, which we set as the average $L_{\rm{IR}}$ of all sources in that particular bin (see Table \ref{tab:dust_prop}).  Once all photometric points are properly redshifted and normalized, we bin the data into wavelength bins from 0.1 $\mu$m to 1000 $\mu$m, with a logarithmic width of 0.05 dex.  We only include bins that have at least 20 data points and where the standard error of the mean ($\sigma/\sqrt{N}$) is smaller than 0.05.  These limits are empirically chosen with the intent of having as much wavelength coverage as possible, but not including erroneous median values that were affected by small number statistics.  Including median values of wavelength bins that don't meet this criteria introduce large, unphysical variations that skew the SEDs.  Figure \ref{fig:all_sed} displays all of the normalized and rest-frame SEDs, with our calculated median SED over plotted.  Error bars for each median point are calculated by measuring the value $\sqrt{N}$ ranks away from the median value, where $N$ is the number of sources in that particular bin.  

\begin{figure*}[!h]
\centering
\includegraphics[width=10 cm]{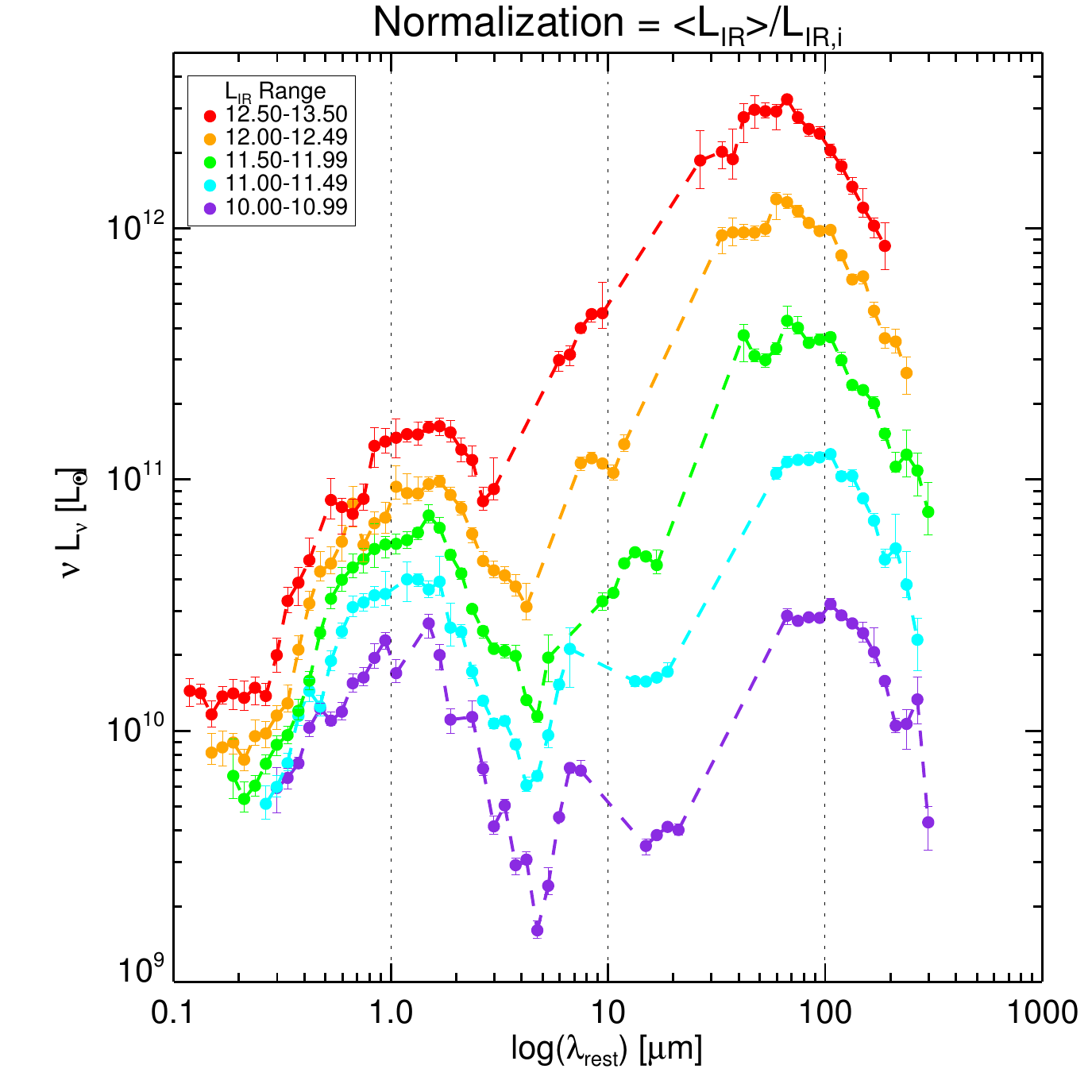}
\caption{Median SEDs from each infrared luminosity bin, all normalized to the average $L_{\rm{IR}}$ of the galaxies in that particular bin.  The median SEDs are the same as those shown in Figure \ref{fig:all_sed}.  Error bars are computed by measuring the value $\sqrt{N}$ ranks above and below the median value, where $N$ is the number of sources in that particular bin.  Vertical lines are plotted to aid the eye in comparing the SEDs.}
\label{fig:med_sed}
\end{figure*}

\begin{figure*}[!h]
\centering
\includegraphics[width=10 cm]{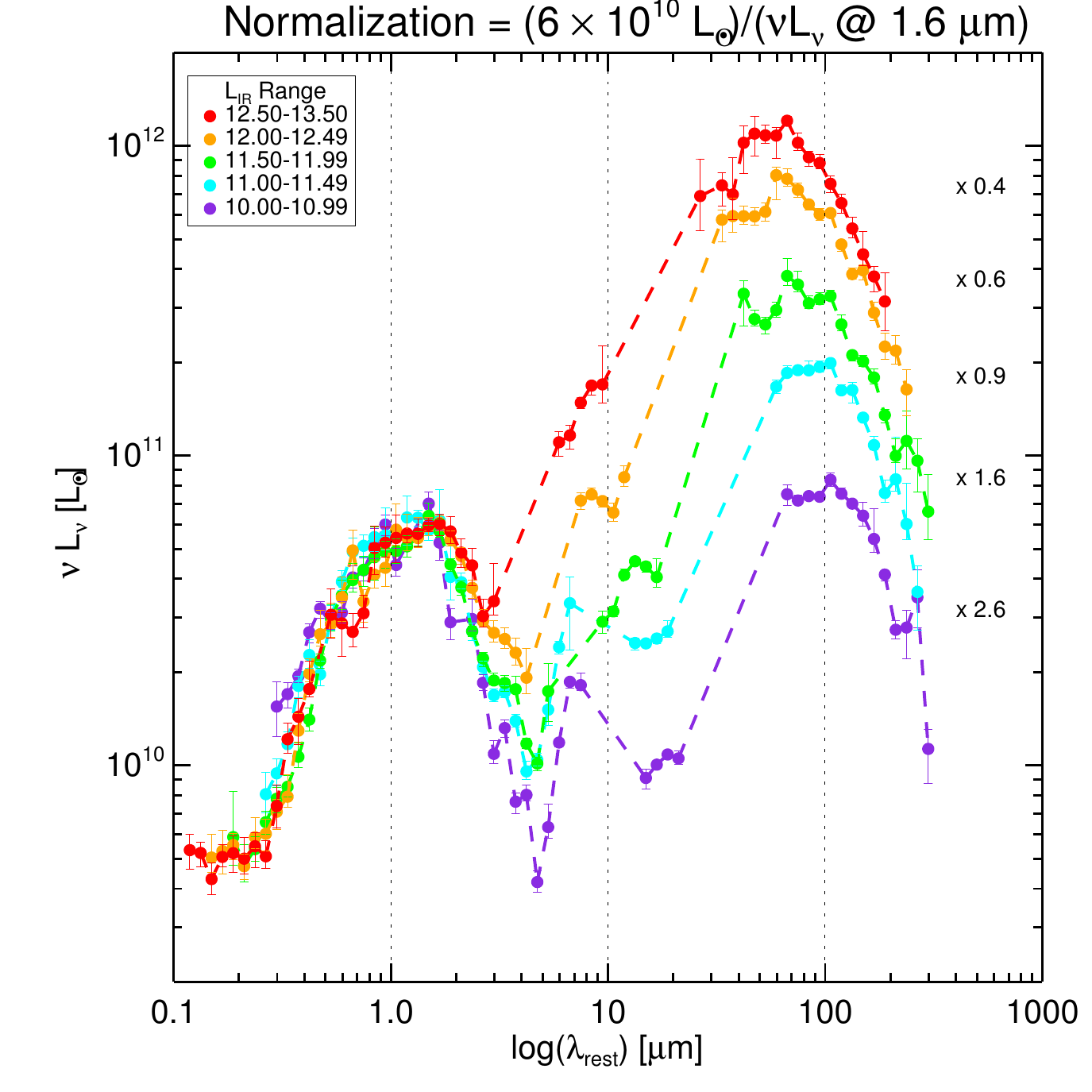}
\caption{Median SEDs from each infrared luminosity bin, all normalized so that $\nu L_{\nu}(1.6 \mu$m)~=~$6 \times 10^{10} L_{\odot}$.  The factor that each SED was multiplied by (when compared to Figure \ref{fig:med_sed}) is shown on the right side of the plot.  This plot emphasizes the differences in emission at both mid-infrared and far-infrared wavelengths between galaxies in different luminosity bins. }
\label{fig:med_sed_1um}
\end{figure*}

\subsection{Median SEDs as a Function of $L_{\rm{IR}}$} \label{sec:sed_evolution}

We plot all the median SEDs together in Figure \ref{fig:med_sed} in order to directly compare the SEDs and see how the typical SED changes with $L_{\rm{IR}}$.  Since the photometry in each bin has been normalized to the median $L_{\rm{IR}}$ in that bin, we see a very clear variation of the strength of the far-infrared emission in each median SED.  Another clear trend is that the far-infrared SEDs peak at shorter wavelengths at higher $L_{\rm{IR}}$, peaking at $\lambda \sim 100 \mu$m in the lowest luminosity bin and peaking at $\lambda \sim 60 \mu$m in the highest luminosity bin.  Since the infrared SED is dominated by blackbody emission at these wavelengths, this suggests that galaxies with higher infrared luminosities have warmer dust temperatures, a trend we also see from our individual galaxies (see Figure \ref{fig:selection_func}).  To better quantify this evolution, we fit each median SED in $S_{\nu}$ using the C12 greybody fits and determine the average dust properties of these galaxies - infrared luminosity, peak wavelength, and dust mass.  The results are listed in Table \ref{tab:dust_prop}, and there is a clear trend in both dust mass and dust temperature toward hotter and more massive dust reservoirs with increasing infrared luminosity.  

We also see the MIR ($\lambda_{\rm{rest}} \sim 10$--$20 \mu$m) portion of the SED increasing with $L_{\rm{IR}}$, suggesting a stronger contribution from warm dust in the form of a MIR power-law.  MIR power-law emission is generally thought to be dominated by emission from dust grains heated to extremely high temperatures ($\approx 1500$--2000$^{\circ}$K) in the dusty tori around AGN, and indeed we even see direct evidence of AGN in the SEDs of a non-negligible fraction of individual galaxies at the highest luminosities.  In Figure \ref{fig:all_sed}, the two highest infrared luminosity bins have some objects with data points at rest-frame UV wavelengths suggestive of a ``big blue bump'' commonly associated with AGN, and thought to be the thermal emission from an optically thick accretion disk surrounding a massive black hole \citep[e.g.][]{1978Natur.272..706S,1982ApJ...254...22M}.  However, we note that these galaxies with putative AGN signatures still make up only a relatively small percentage of the sample and removing galaxies classified as AGN does not affect our median SEDs significantly. 

\begin{deluxetable*}{ccccccc}
\tablewidth{0pt}
\tablecaption{Average Properties of {\em Herschel}-selected galaxies \label{tab:dust_prop}}
\tablehead{\colhead{log$(L_{\rm{IR}})$ Range} & \colhead{$\langle z \rangle^{a}$} & \colhead{$\langle$log$(M_{*})\rangle^{b}$} & \colhead{log$(L_{\rm{IR}})^{c}$} &  \colhead{log$(M_{\rm{dust}})^{d}$} & \colhead{$\lambda_{\rm{peak}}$$^{e}$} & \colhead{$T_{\rm{dust,grey}}$$^{f}$} \\
\colhead{[$L_{\odot}$]} & \colhead{} & \colhead{[$M_{\odot}$]} & \colhead{[$L_{\odot}$]} & \colhead{[$M_{\odot}$]} & \colhead{[$\mu$m]} & \colhead{[K]}  }
\startdata
10.00--10.99 & 0.30 & 10.4 & 10.8 &  7.6 & 140 & 27.8  \\
11.00--11.49 & 0.56 & 10.5 & 11.4 &  8.0 & 124 & 32.8  \\
11.50--11.99 & 0.93 & 10.6 & 11.9 &  8.4 & 112 & 37.5  \\
12.00--12.49 & 1.46 & 10.8 & 12.4 &  8.4 &  86 & 53.2  \\
12.50--13.49 & 2.19 & 10.9 & 12.8 &  8.8 &  85 & 54.1  

\enddata
\tablecomments{$(a)$ Median photo-z of all sources in $L_{\rm{IR}}$ bin.  $(b)$ Median stellar mass of all sources in $L_{\rm{IR}}$ bin from \citet[][see \S\ref{sec:mass_sfr}]{2010ApJ...709..644I}. $(c)$ $L_{\rm{IR}}$ measured from C12 fits to median SED.  $(d)$ Dust mass measured from C12 fit to median SED. $(e)$ Rest-frame SED peak wavelength of $S_{\nu}$ from C12 fit to median SEDs. $(f)$ Dust temperature assuming a grey-body with $\beta = 1.5$ and $\tau = 1$ at 200~$\mu$m, derived from $\lambda_{\rm{peak}}$ of median SEDs.  }
\end{deluxetable*}

Although we do not have the uniform wavelength coverage to make detailed comparisons across all luminosity bins, we also see possible PAH signatures at MIR wavelengths.  In the two lowest luminosity bins, we see evidence of an emission feature at $\lambda_{\rm{rest}} \approx 7$--$8 \mu$m, presumably due to the strong PAH emission line at $7.71 \mu$m \citep{1999ESASP.427..579T}.  The SEDs for the two highest infrared luminosity bins also show hints of an absorption feature at $\lambda_{\rm{rest}} \approx 10 \mu$m that could possibly be explained as silicate absorption.  We explore the variation of MIR features (specifically at $\lambda_{\rm{rest}} \approx 8 \mu$m) with $L_{\rm{IR}}$ in more detail in \S\ref{sec:ir8}.

An additional trend seen at shorter wavelengths is that at increasing infrared luminosities, the SEDs also have increased luminosity at optical and near-infrared (NIR: $\lambda_{\rm{rest}} \approx 2$--$5 \mu$m) wavelengths.  However, while the average infrared luminosities across our bins increase by a factor of $\sim 100$, the optical-NIR luminosities only increase by a factor of $\sim 10$.  Near-infrared emission comes mostly from relatively older, cooler, and less massive stars that dominate the stellar mass of a galaxy, so this suggests that our more infrared luminous sources have larger stellar masses.  To investigate this further, we normalize all of the median SEDs at $1.6 \mu$m to an arbitrary value of $\nu L_{\nu}(1.6 \mu$m$) = 10^{9} L_{\odot}$ and plot the result in Figure \ref{fig:med_sed_1um}.  The relative difference between far-infrared luminosity and optical-NIR luminosity (or stellar mass) is clearly displayed, with the highest $L_{\rm{IR}}$ bins showing more than an order of magnitude difference between FIR luminosity and optical-NIR luminosity, while the lowest $L_{\rm{IR}}$ bin has almost equal FIR luminosity and optical-NIR luminosity.  We explore the relationship between stellar mass ($M_{*}$) and $L_{\rm{IR}}$ in \S\ref{sec:mass_sfr}.

\subsection{Trends with Redshift or Luminosity?}\label{sec:selection_effects}

A major issue we must reconcile when studying our median SEDs is the degeneracy between redshift and infrared luminosity.  As seen in Figure \ref{fig:lir_casey_vs_ce}, redshift and infrared luminosity are correlated.  This is because at low redshift, we do not cover enough volume to properly sample the high luminosity population, and at high redshift, the low luminosity galaxies fall below our detection limits.  In addition, at low redshifts, the number density of high infrared luminosity galaxies drop dramatically, which makes detection of the highest luminosity galaxies difficult in the lowest redshift bins.  Thus, it is possible that instead of witnessing evolution of SEDs with infrared luminosity, we are simply looking at galaxies at different redshifts that have evolving SEDs because of redshift evolution.  

We tested if the trends with luminosity seen in \S\ref{sec:sed_evolution} were actually due to redshift evolution by splitting each luminosity bin sample into redshift bins with width $\Delta z = 0.5$, and then constructing new median SEDs with the smaller subsamples.  For every luminosity bin, we find that the new median SEDs in all redshift slices were consistent with each other, and did not evolve significantly with redshift.  This suggests that the main cause of the variation in SEDs is indeed the infrared luminosity of the objects.  

\section{Discussion}\label{sec:discussion}

Our large sample of 4,218 {\em Herschel}-selected galaxies in the COSMOS field has permitted us to determine basic properties of luminous infrared galaxies at redshifts out to $z \sim 3.5$.  Although this is not the first large sample of (U)LIRGs at high redshift, our analysis is the first to systematically investigate the median SED properties of (U)LIRGs as a function of $L_{\rm{IR}}$ at these redshifts.  One of the more surprising results from previous studies has been the suggestion that the majority of luminous infrared galaxies at high redshift form stars in a ``normal main sequence mode''.  \citet{2007ApJ...660L..43N} find a tight correlation between stellar mass ($M_{*}$) and star formation rate (SFR), and that this entire correlation shifts towards lower specific star formation rates ($sSFR \equiv SFR/M_{*}$) with time by a factor of $\sim 3$ from $z = 0.98$ to $z = 0.36$.  \citet[][hereafter E11]{2011A&A...533A.119E} find that infrared luminous galaxies seem to fall on an ``infrared main sequence'' with a constant value of IR8, defined as the ratio of total infrared luminosity ($L_{\rm{IR}}$) to luminosity at restframe $8 \mu$m ($L_{8} \equiv \nu L_{\nu}(8 \mu$m)).  

At low redshifts, both of these ``main sequence'' relations appear to hold for galaxies with $L_{\rm{IR}} < 10^{11.3} L_{\odot}$, but at higher infrared luminosities, galaxies systematically lie above both the $SFR/M_{*}$ ``main-sequence'' and the IR8 ``infrared main sequence''.  At higher redshifts, these ``main sequences'' appear to shift to higher SFRs (or $L_{\rm{IR}}$), which means that LIRGs and ULIRGs begin to overlap with the main sequence, begging the question - are (U)LIRGs at high redshift simply scaled up versions of lower luminosity galaxies?  This has many important implications for understanding star-formation at high redshift, and we wish to investigate these questions using our new large sample of objects.  

\subsection{Stellar Mass and Star Formation Rate}\label{sec:mass_sfr}

\begin{figure}[!h]
\centering
\includegraphics[width=8 cm]{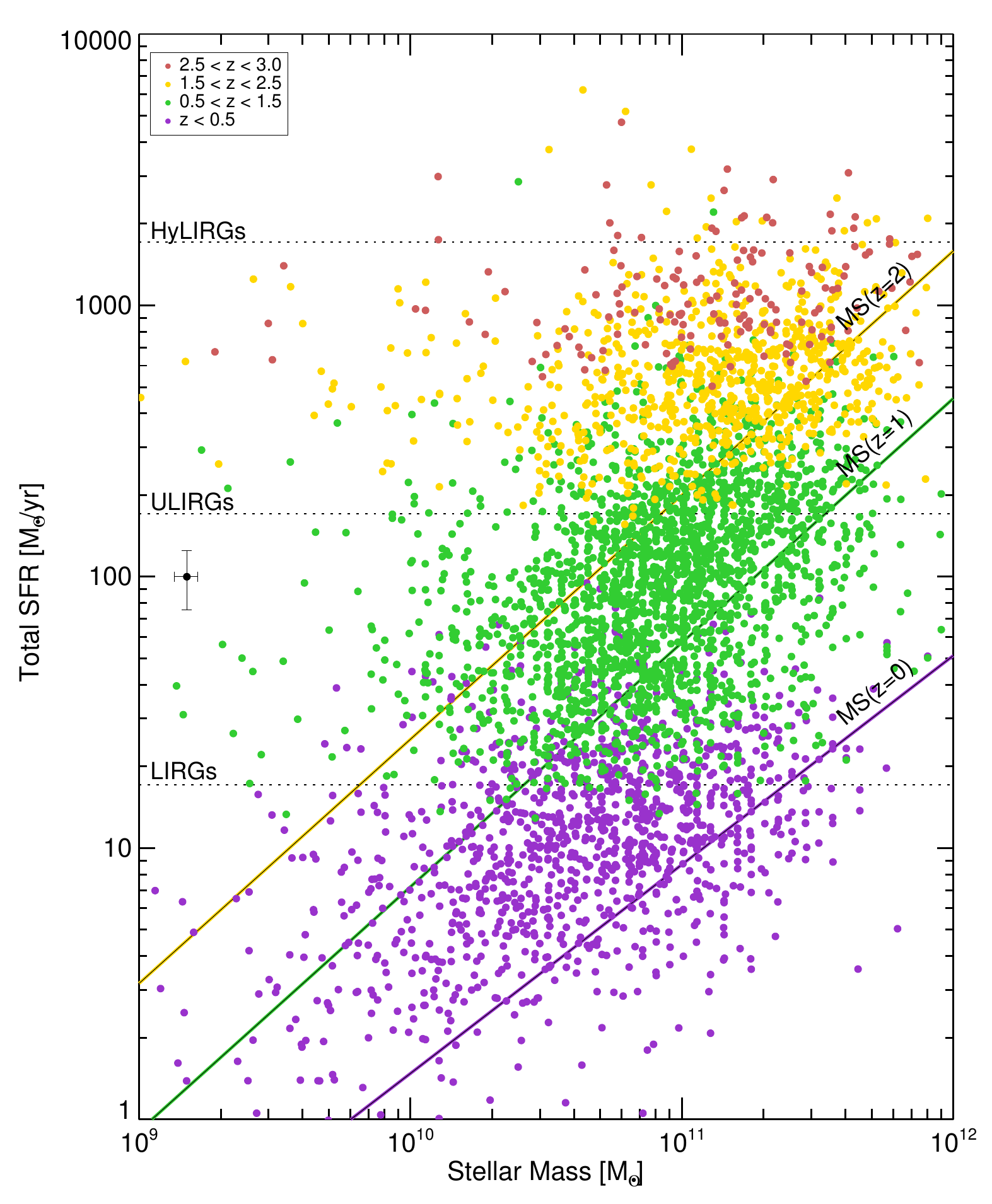}
\caption{Stellar Mass vs. SFR of our entire {\em Herschel} sample, colored by redshift.  Our redshift bins span $z < 0.5$ (purple), $0.5 < z < 1.5$ (blue), $1.5 < z < 2.5$ (yellow), and $2.5 < z < 3.0$ (red).  Colored solid lines represent the ``main-sequence'' of galaxies at $z \sim 0$ \citep[purple,][]{2007ApJ...670..156D}, $z \sim 1$  \citep[blue,][]{2007ApJ...670..156D}, and $z \sim 2$ \citep[yellow,][]{2007A&A...468...33E}, and the horizontal dotted lines simply mark the star formation rates that correspond to the infrared luminosities of LIRGs, ULIRGs, and HyLIRGs.}
\label{fig:mass_lir}
\end{figure}

To investigate the relationship between $M_{*}$ and SFR for our complete sample of {\em Herschel}-selected galaxies, we use stellar masses from \citet{2010ApJ...709..644I}, who use stellar population synthesis models from \citet{2003MNRAS.344.1000B} with an IMF from \citet{2003PASP..115..763C} and an exponentially declining star formation history, and we convert these masses to a \citet{1955ApJ...121..161S} IMF by multiplying by a factor of 1.8 \citep[as in][]{2012A&A...541A..85M}.  We calculate total SFRs for each galaxy by combining their unobscured SFR from GALEX \citep{2013ApJS..206....8M} with obscured SFR from $L_{\rm{IR}}$ using the conversion 
\begin{equation}
SFR(M_{\odot}~yr^{-1}) = 4.5 \times 10^{-44} L_{\rm{IR}} (erg~s^{-1})
\end{equation} 
given in \citet{1998ARA&A..36..189K}, although due to our {\em Herschel} selection, the obscured star formation dominates the SFR in most of our galaxies.  We plot SFR and $M_{*}$ of all our galaxies in Figure \ref{fig:mass_lir}.  For comparison, we also include ``main sequence'' lines for three redshift bins: $z \sim 0, 1, 2$ \citep{2007ApJ...670..156D,2007A&A...468...33E}.    

At all redshifts, we see no evidence that the {\em Herschel}-detected galaxies in the COSMOS field concentrate on the nominal main-sequence trends plotted, but instead appear to have a much flatter distribution with stellar mass.  We stress that {\em Herschel} observations of COSMOS generally sample only the most luminous regime of the SFR/$M_{*}$ plane, so we can not make any statements about galaxies at lower luminosities.  Indeed, in a mass-selected sample, the {\em Herschel}-selected galaxies represent only a small percentage of the total number of galaxies \citep[e.g.][]{2011ApJ...739L..40R}.  However, our sensitivity in SFR is sufficient enough such that we should still see evidence of galaxies clustering or concentrating around the ``main sequence'' relation if such were the case.  To demonstrate this, we model a population of 4000 galaxies at a redshift $z = 1$ with masses between $9.5 < $log($M_{*}) < 11.5$ that follow the SFR/$M_{*}$ main-sequence from \citet{2010ApJ...718.1001B}, who provide a redshift-dependent functional form of the main-sequence.  We then simulate our {\em Herschel} selection by removing all sources with SFR$< 50 M_{*}$/yr, the approximate selection function at $z = 1$ (see Figure \ref{fig:selection_func}).  Over 1000 such simulations, we find an average of 37\% of the remaining simulated galaxies lie more than 0.3 dex off the main-sequence, with a maximum of 44\%.  In contrast, $\sim 60$\% our {\em Herschel} galaxies lie more than 0.3 dex above the redshift dependent main-sequence.  Thus, we find almost twice as many galaxies above the SFR/$M_{*}$ main-sequence as would be expected.

These results are similar to those found in \citet{2011ApJ...739L..40R}, who see that the {\em Herschel} PACS detected galaxies in their sample have a relatively flat distribution across stellar mass.  However, \citet{2011ApJ...739L..40R} suggest that the presence of a large population of color-selected $BzK$ galaxies that lie on the main-sequence dominate the number counts such that the overall population of star-forming galaxies still follows the general main-sequence trend.  This may be the case at low masses, but at the high mass end where the main-sequence lies above the {\em Herschel} sensitivity limits, this cannot be the case unless there exists a large population of extremely UV bright, infrared dim objects with high star formation rates ($SFR > 100~M_{\odot}/yr$) that are missed by {\em Herschel}.  A more likely explanation is that SFR indicators based on IR data, which directly measure the obscured star formation, differ greatly from SFR indicators that are based on measuring the unobscured star formation rate and applying a correction for dust.  Indeed, for our sample of {\em Herschel} sources, we find that the optically derived total SFRs (corrected for dust) underpredict the far-infrared derived total SFR by a factor of 2.7 (median) to 9.6 (mean). 

Because our sample is essentially a SFR selection, we cannot infer anything about the main-sequence below our selection limits. It may be that galaxies at low star formation rates follow the ``main sequence'', but at star formation rates above a specific limit (e.g. SFR$ > 100 M_{\odot}/yr$ at $z \sim 1$), galaxies deviate significantly from the main sequence, as seen in local galaxy samples (Larson et al. in prep).  One possible physical explanation may be that the high star formation rates probed by {\em Herschel} observations require more extreme physical processes, such as galaxy mergers \citep{2013arXiv1309.4459H}.

\subsection{Mid-Infrared to Far-Infrared Diagnostics}\label{sec:ir8}

Recent studies of {\em Herschel}-selected galaxies in GOODS-N and GOODS-S \citep[E11,][]{2012ApJ...745..182N} have concluded that most infrared luminous galaxies at redshifts $z \sim 0$--3 have a constant ratio of total infrared luminosity ($L_{\rm{IR}}$) to $\nu L_{\nu}$ at $8 \mu$m ($L_{8}$), defined as IR8~$\equiv L_{\rm{IR}} / L_{8}$.  E11 find that most infrared luminous galaxies at these redshifts follow a Gaussian distribution centered on IR8 = 4 ($\sigma = 1.6$), which they claim defines an ``infrared main sequence for star-forming galaxies independent of redshift and luminosity''. Those few galaxies which lie above the ``infrared main sequence'' were classified by E11 as a population of ``compact starburst galaxies'', as opposed to more extended star-forming regions which were assumed to be representative of the larger population of galaxies on the ``infrared main sequence''.  E11 note that these new results were contrary to what is observed in samples of local infrared luminous galaxies, which show a constant IR8 value at infrared luminosities below $L_{\rm{IR}} < 10^{11} L_{\odot}$, but have a systematic increase in IR8 at higher luminosities (see Figure 8 in E11).  These results have contributed to the suggestion that the large majority of (U)LIRGs at high redshift form stars in a ``normal main sequence'' mode, as opposed to more local (U)LIRGs.  We wish to test this important new result using our large sample of high-redshift infrared luminous galaxies.

As discussed in \S\ref{sec:sed_evolution}, we see hints that the MIR PAH features in the median SEDs of the {\em Herschel}-selected galaxies in the COSMOS field do in fact vary with infrared luminosity.  However, our wavelength coverage is not sufficient to draw meaningful conclusions from the median SEDs alone.  In order to compare our studies more directly, we calculate IR8 for each source in our sample.  Since we do not have a direct measure of $\nu L_{\nu}$ at $8 \mu$m (rest frame) for all of our objects, we extrapolate from observed {\em Spitzer} MIPS $24 \mu$m or IRAC $8 \mu$m fluxes by assuming an SED shape.  E11 used an M82 SED template for all of their extrapolations, but had additional coverage at $16 \mu$m from {\em Spitzer} IRS peak-up array imaging, in addition to {\em Spitzer} MIPS $24 \mu$m or IRAC $8 \mu$m observations, which means they required less extrapolations around $z \sim 1$.  By measuring $L_{8}$ using extrapolations from both $16 \mu$m and $24 \mu$m for GOODS galaxies with observations at both wavelengths, we find that calculating $L_{8}$ from $24 \mu$m around $z \sim 1$ (as we do in COSMOS) generally matches the extrapolations from $16 \mu$m, with a scatter of a factor of $\sim 2$ and no systematic offset (private communication D. Elbaz).  

\begin{figure}[!h]
\centering
\includegraphics[width=8 cm]{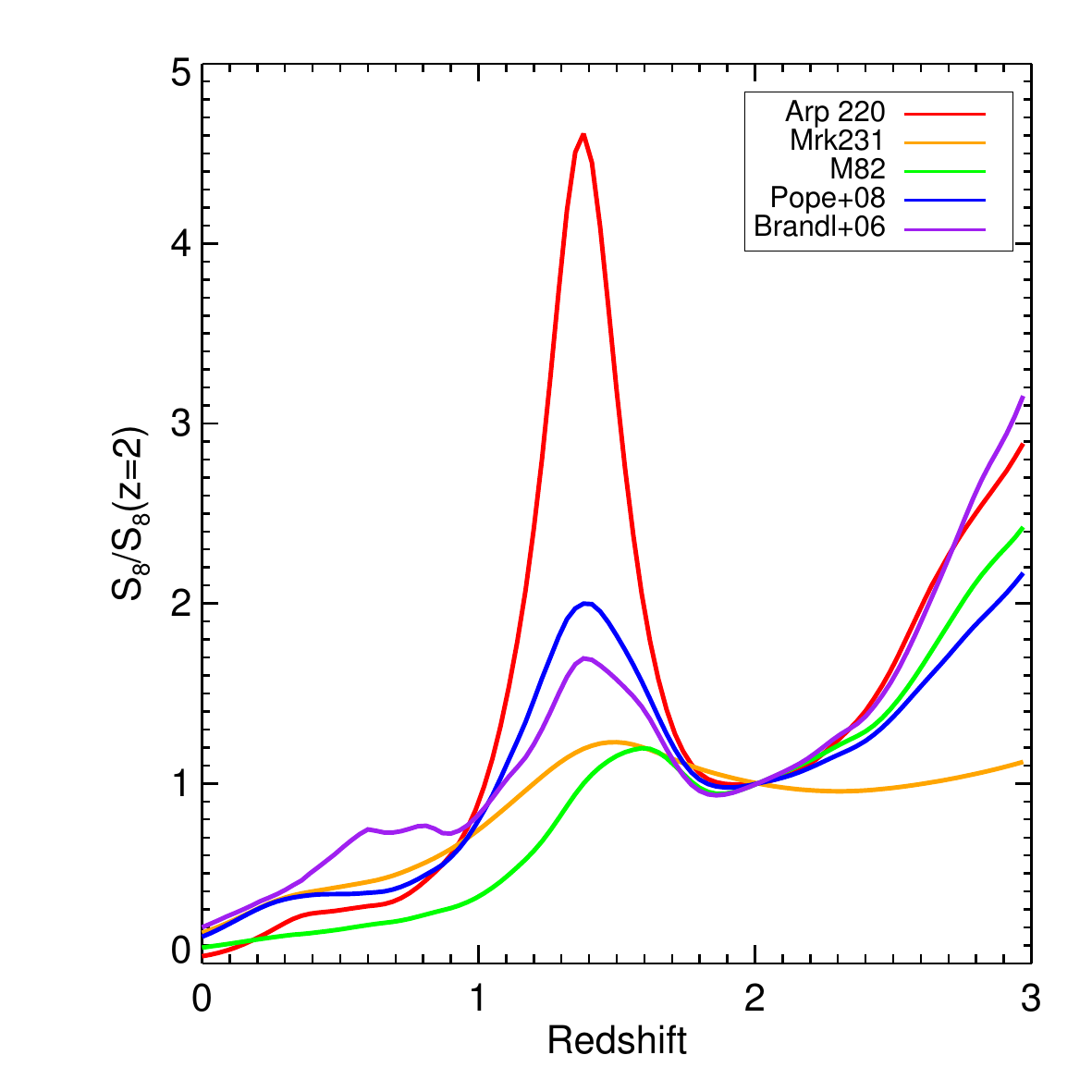}
\caption{The effects of SED template choice on correction factors when calculating $L_{8}~=~\nu L_{\nu}(8 \mu$m).  The different colors correspond to different templates that were used: Arp 220 (red), Mrk 231 (orange) and M82 (green) from the SWIRE template library \citep{2007ApJ...663...81P}, mid-IR composite spectra from 13 SMGs from \citet[][blue]{2008ApJ...675.1171P}, and mid-IR composite spectra from starburst galaxies from \citet[][purple]{2006ApJ...653.1129B}.  The value plotted ($S_{8}/S_{8}(z=2)$), gives the ratio of the correction factor applied at the desired redshift to the factor applied at a redshift of $z=2$ (notice that every line crosses unity at $z=2$).  We compare to $z=2$ because we have $24 \mu$m observations for each source, and at $z=2$, the $24 \mu$m directly probes rest frame $8 \mu$m and so has limited correction factors anyway (slight corrections are needed to account for the different passbands in MIPS $24 \mu$m and IRAC $8 \mu$m).  For redshifts below $z = 1$, we calculate corrections using the observed IRAC $8 \mu$m flux densities, while at all other redshifts we use the MIPS $24 \mu$m flux densities.  This plot displays the dangers in assuming a single ``star-forming'' SED template to calculate $L_{8}$, as the template you choose can drastically affect the inferred luminosities.}
\label{fig:l8_temp}
\end{figure}

Although a single SED template from M82 was used in E11, we were concerned about assuming a single SED template because rest-frame $8 \mu$m lies in a forest of PAH features, and the choice of model used for extrapolation could affect the results drastically.  We demonstrate this in Figure \ref{fig:l8_temp}, where we plot the extrapolation factors at different redshifts when using SED templates from M82, Mrk231, Arp220 \citep{2007ApJ...663...81P}, and star-formation SEDs from \citet{2006ApJ...653.1129B} and \citet{2008ApJ...675.1171P}.  There can be a large variation in derived $L_{8}$ depending on which SED template is used.  For the remainder of our analysis, we calculate $L_{8}$ for our galaxies by using the average $L_{8}$ calculated from the models of star-forming galaxies \citep[][and M82]{2006ApJ...653.1129B,2008ApJ...675.1171P}, and using the standard deviation as an additional error in $L_{8}$.   We also repeat all analyses using each single SED template in Figure \ref{fig:l8_temp} to see if the use of a particular template affects the results.  We find that while the choice of template can affect broad changes (particularly around $z \sim 1.5$ due to fitting of the strong $10 \mu$m absorption feature), the overall trends we discuss do not differ significantly with the choice of model.  

\begin{figure*}[!h]
\centering
\includegraphics[width=12 cm]{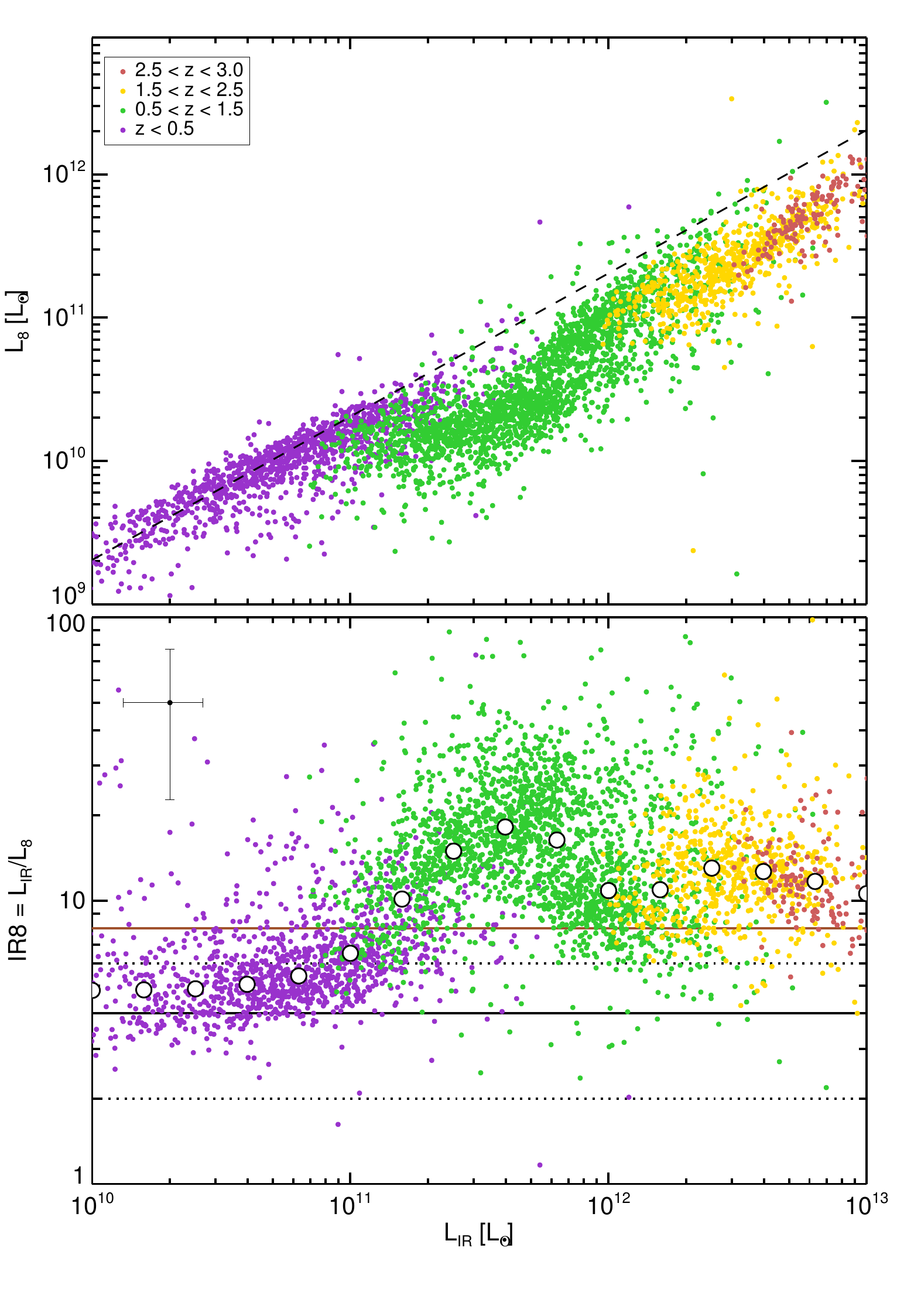}
\caption{{\em Top.} A plot comparing $L_{\rm{IR}}$ and $L_{8}$ for our {\em Herschel} COSMOS sources, with each symbol colored by redshift.  The dashed line represents the relationship quoted in E11.  We see that the majority of our detected sources fall off the E11 relationship, in general having much higher $L_{\rm{IR}}$ than expected from the E11 relationship.  {\em Bottom.}  A plot detailing how IR8 ($\equiv L_{\rm{IR}} / L_{8}$) changes with infrared luminosity, with colored dots representing the same sources as in the top plot.  A running median of IR8 in bins of $L_{\rm{IR}}$ is displayed with white circles.  The black solid line represents the IR8 main-sequence defined by E11, with upper and lower limits drawn as black dotted lines.  The maroon horizontal line represents the lower bound for galaxies deemed as ``starburst'' from their IR8.  Typical uncertainties are plotted in the upper left hand corner.}
\label{fig:ir8}
\end{figure*}

The results of our analysis are plotted in Figure \ref{fig:ir8}, with the top plot displaying how $L_{8}$ varies with $L_{\rm{IR}}$, and the bottom plot showing how IR8 varies with $L_{\rm{IR}}$.  In both plots, we have included a line displaying the ``main sequence'' from E11, and we see that our sources do not follow a single main sequence, as seen in E11, but instead seem to scatter to much higher IR8 values at infrared luminosities $L_{\rm{IR}} \gtrsim 10^{11} L_{\odot}$, similar to what is seen in the local universe.

The trend we see in IR8 for {\em Herschel}-selected galaxies in COSMOS is actually very similar to the trend seen in local galaxies, where galaxies below $L_{\rm{IR}} \sim 10^{11} L_{\odot}$ lie near the IR8 ``infrared main sequence'' and galaxies with higher infrared luminosities show a systematic increase in the value of IR8 vs. $L_{\rm{IR}}$ (see Fig. 8 in E11).  Could our results be due to a lack of depth in observations?  When compared to the observations from the Great Observatories Origins Deep Survey fields (GOODS, E11) that the original ``IR8 main-sequence'' was based on, we see that we have shallower coverage at both 24 $\mu$m and in {\em Herschel}, but with much larger volume.  It may be that sources with $L_{\rm{IR}}$ below our far-infrared detection threshold will fall closer to the ``IR8 main-sequence,'' but at each redshift, these galaxies will necessarily fall to the left of the locus of points seen in the bottom panel of Figure \ref{fig:ir8}, and will not be able to change the median location of those galaxies.  It may also be possible that we are missing galaxies with similar $L_{\rm{IR}}$ to our sample, but were not detected because we require a 24 $\mu$m prior, and these galaxies have depressed $S_{24}$.  However, these galaxies must necessarily have high IR8($=L_{\rm{IR}}/$L8) values, and will push our results even farther from the ``IR8 main-sequence''.

It appears that galaxies at high redshift follow the same general trend as the local galaxies, with lower luminosity galaxies falling along the ``IR8 main-sequence'' and higher luminosity galaxies scattering to higher IR8, but the cutoff luminosity increases with redshift, due to the increasing gas density and star formation at earlier epochs in the Universe.  This interpretation might also explain the discrepancy between our results and those of E11.  GOODS ($\sim 260$ arcmin$^{2}$) is a much smaller field than COSMOS ($\sim 2$ deg$^{2}$), but has PACS observations that are $\sim 3$ times deeper than those of the COSMOS field.  As a result of these discrepancies, GOODS observations are more sensitive to low luminosity galaxies, but do not have the volume coverage to find the rare, luminous sources, whereas {\em Herschel}-COSMOS observations cannot detect the lowest luminosity galaxies.  Since the GOODS observations did not have enough volume to probe the high luminosity (high IR8) galaxies at any redshift, E11 find a ``continuous'' main sequence by combining the lower luminosity galaxies in each redshift slice.  Although the {\em Herschel}-COSMOS observations do not have the sensitivity required to fully sample the lower luminosity galaxies that may lie on the IR8 ``infrared main sequence,'' there is sufficient volume coverage to better sample the high infrared luminosity population.  

It seems that at the infrared luminosities probed by {\em Herschel}-COSMOS, the majority of luminous infrared sources at all redshifts do not follow either the $SFR/M_{*}$ ``main sequence'', nor the IR8 ``infrared main sequence''.  In general, we see similar trends with $L_{\rm{IR}}$ as observed for (U)LIRGs in the local universe, where galaxies with $L_{\rm{IR}} \lesssim 10^{11} L_{\odot}$ lie on these ``main sequences'' while higher luminosity galaxies lie above the ``main sequences''.

\section{Conclusions}

We have used new {\em Herschel} PACS \& SPIRE observations of the large, contiguous 2 deg$^{2}$ COSMOS field in order to identify and study the multi-wavelength properties of 4,218 infrared luminous galaxies.  {\em Spitzer} $24 \mu$m counterparts were used to match our {\em Herschel} sources to existing multi-wavelength photometry, allowing us to construct full rest-frame UV-to-FIR SEDs and determine accurate photometric redshifts.  Our sources span a redshift range of $0.02 < z < 3.54$ and a total infrared luminosity range of log$(L_{\rm{IR}}/L_{\odot}) = 9.4$--13.6.  We determine the basic properties of each galaxy (e.g. $L_{\rm{IR}}$, $M_{\rm{dust}}$, and $\lambda_{\rm{peak}}$) by fitting their infrared SEDs to a coupled modified greybody plus a MIR power law.  In order to study the galaxy SEDs in more detail, we then compute median SEDs, binned by their total infrared luminosities.  

From our detailed analysis of the COSMOS {\em Herschel}-selected galaxies, we find the following major results:

\begin{enumerate}
\item The SED peak wavelength systematically decreases from $\lambda_{\rm{peak}} \sim 140 \mu$m at $L_{\rm{IR}} \sim 10^{10.8} L_{\odot}$ to $85 \mu$m at $L_{\rm{IR}} \sim 10^{12.8} L_{\odot}$.  Over the same luminosity range, the dust mass systematically increases from log($M_{\rm{dust}}/M_{\odot}) = 7.6$ to 8.8.
\item A comparison of the average luminosities at FIR wavelengths and at optical-NIR wavelengths shows that as  $L_{\rm{IR}}$ increases by a factor of 100 (from $L_{\rm{IR}} = 10.8$--12.8), the stellar mass increases by only a factor of $\sim 3$ (from log$(M_{*}/M_{\odot}) = 10.4$--10.9).
\item At lower infrared luminosities ($L_{\rm{IR}} < 11.5$), we see evidence of PAH features in the MIR ($\lambda_{\rm{rest}} \approx 8 \mu$m) which appear less significant at higher luminosities, where we see an apparent increasing contribution of hot dust ($T_{\rm{d}} \sim 100$--300$^{\circ}$K) corresponding to the emergence of a power-law component at $\lambda_{\rm{rest}} \approx 3$--$30 \mu$m.  At the highest luminosities ($L_{\rm{IR}} > 10^{12} L_{\odot}$), we see a small, but increasing fraction of objects with prominent UV \& optical excess, similar to the ``big blue bump'' seen in optically-selected QSOs.
\item We find no evidence that our {\em Herschel}-selected luminous infrared galaxies in COSMOS lie on the $SFR/M_{*}$ ``main sequence'' previously defined by studies of optically selected galaxies.  About 60\% of our luminous infrared galaxies lie more than 1$\sigma$ above the ``main sequence'' relationship.
\item We find no evidence that a constant value of IR8 ($\equiv L_{\rm{IR}} / L_{8}$) applies to infrared luminous galaxies at high redshift.  Instead, we find that at low infrared luminosities, galaxies have a constant value of IR8 ($\approx 4\pm1.6$), but at high infrared luminosities ($L_{\rm{IR}} \gtrsim 10^{11.3} L_{\odot}$), galaxies systematically lie above the IR8 ``infrared main sequence'', similar to what is seen for (U)LIRGs in the local universe.
\end{enumerate} 

This is the first in a series of papers that will explore in more detail the properties of infrared luminous galaxies across cosmic time.  Future papers will study morphologies, spectral types, comparisons of UV/optical and far-IR derived SFRs, and comparisons of the high redshift population with complete samples of (U)LIRGs in the local universe ($z < 0.3$).

\section{Acknowledgements}
D. B. Sanders and C. M. Casey acknowledge the hospitality of the Aspen
Center for Physics, which is supported by the National Science
Foundation Grant No. PHY-1066293.
C. M. Casey is generously supported by a Hubble Fellowship from Space
Telescope Science Institute, grant HST-HF-51268.01-A.

COSMOS is based on observations with the NASA/ESA {\em Hubble Space
Telescope}, obtained at the Space Telescope Science Institute, which
is operated by AURA Inc, under NASA contract NAS 5-26555; also based
on data collected at : the Subaru Telescope, which is operated by the
National Astronomical Observatory of Japan; the XMM-Newton, an ESA
science mission with instruments and contributions directly funded by
ESA Member States and NASA; the European Southern Observatory, Chile;
Kitt Peak National Observatory, Cerro Tololo Inter-American
Observatory, and the National Optical Astronomy Observatory, which are
operated by the Association of Universities for Research in Astronomy,
Inc. (AURA) under cooperative agreement with the National Science
Foundation; the National Radio Astronomy Observatory which is a
facility of the National Science Foundation operated under cooperative
agreement by Associated Universities, Inc ; and the
Canada-France-Hawaii Telescope operated by the National Research
Council of Canada, the Centre National de la Recherche Scientifique de
France and the University of Hawaii.

PACS has been developed by a consortium of institutes led by MPE
(Germany) and including UVIE (Austria); KU Leuven, CSL, IMEC
(Belgium); CEA, LAM (France); MPIA (Germany); INAF-IFSI/OAA/OAP/OAT,
LENS, SISSA (Italy); IAC (Spain). This development has been supported
by the funding agencies BMVIT (Austria), ESA-PRODEX (Belgium),
CEA/CNES (France), DLR (Germany), ASI/INAF (Italy), and CICYT/MCYT
(Spain).

SPIRE has been developed by a consortium of institutes led by Cardiff
University (UK) and including Univ. Lethbridge (Canada); NAOC (China);
CEA, LAM (France); IFSI, Univ. Padua (Italy); IAC (Spain); Stockholm
Observatory (Sweden); Imperial College London, RAL, UCL-MSSL, UKATC,
Univ. Sussex (UK); and Caltech, JPL, NHSC, Univ. Colorado (USA). This
development has been supported by national funding agencies: CSA
(Canada); NAOC (China); CEA, CNES, CNRS (France); ASI (Italy); MCINN
(Spain); SNSB (Sweden); STFC (UK); and NASA (USA).

\bibliographystyle{apj}
\bibliography{apj-jour,irseds}

\begin{thebibliography}{67}
\expandafter\ifx\csname natexlab\endcsname\relax\def\natexlab#1{#1}\fi

\bibitem[{{Aretxaga} {et~al.}(2011){Aretxaga}, {Wilson}, {Aguilar}, {Alberts},
  {Scott}, {Scoville}, {Yun}, {Austermann}, {Downes}, {Ezawa}, {Hatsukade},
  {Hughes}, {Kawabe}, {Kohno}, {Oshima}, {Perera}, {Tamura}, \&
  {Zeballos}}]{2011MNRAS.415.3831A}
{Aretxaga}, I., {et~al.} 2011, \mnras, 415, 3831

\bibitem[{{Armus} {et~al.}(2009){Armus}, {Mazzarella}, {Evans}, {Surace},
  {Sanders}, {Iwasawa}, {Frayer}, {Howell}, {Chan}, {Petric}, {Vavilkin},
  {Kim}, {Haan}, {Inami}, {Murphy}, {Appleton}, {Barnes}, {Bothun}, {Bridge},
  {Charmandaris}, {Jensen}, {Kewley}, {Lord}, {Madore}, {Marshall},
  {Melbourne}, {Rich}, {Satyapal}, {Schulz}, {Spoon}, {Sturm}, {U}, {Veilleux},
  \& {Xu}}]{2009PASP..121..559A}
{Armus}, L., {et~al.} 2009, \pasp, 121, 559

\bibitem[{{Bertoldi} {et~al.}(2007){Bertoldi}, {Carilli}, {Aravena},
  {Schinnerer}, {Voss}, {Smolcic}, {Jahnke}, {Scoville}, {Blain}, {Menten},
  {Lutz}, {Brusa}, {Taniguchi}, {Capak}, {Mobasher}, {Lilly}, {Thompson},
  {Aussel}, {Kreysa}, {Hasinger}, {Aguirre}, {Schlaerth}, \&
  {Koekemoer}}]{2007ApJS..172..132B}
{Bertoldi}, F., {et~al.} 2007, \apjs, 172, 132

\bibitem[{{Blain} {et~al.}(2004)}]{2004ApJ...611...52B}
{Blain}, A.~W., {et~al.} 2004, \apj, 611, 52

\bibitem[{{Bouch{\'e}} {et~al.}(2010){Bouch{\'e}}, {Dekel}, {Genzel}, {Genel},
  {Cresci}, {F{\"o}rster Schreiber}, {Shapiro}, {Davies}, \&
  {Tacconi}}]{2010ApJ...718.1001B}
{Bouch{\'e}}, N., {et~al.} 2010, ApJ, 718, 1001

\bibitem[{{Brandl} {et~al.}(2006){Brandl}, {Bernard-Salas}, {Spoon}, {Devost},
  {Sloan}, {Guilles}, {Wu}, {Houck}, {Weedman}, {Armus}, {Appleton}, {Soifer},
  {Charmandaris}, {Hao}, {Higdon}, {Marshall}, \&
  {Herter}}]{2006ApJ...653.1129B}
{Brandl}, B.~R., {et~al.} 2006, \apj, 653, 1129

\bibitem[{{Brusa} {et~al.}(2010){Brusa}, {Civano}, {Comastri}, {Miyaji},
  {Salvato}, {Zamorani}, {Cappelluti}, {Fiore}, {Hasinger}, {Mainieri},
  {Merloni}, {Bongiorno}, {Capak}, {Elvis}, {Gilli}, {Hao}, {Jahnke},
  {Koekemoer}, {Ilbert}, {Le Floc'h}, {Lusso}, {Mignoli}, {Schinnerer},
  {Silverman}, {Treister}, {Trump}, {Vignali}, {Zamojski}, {Aldcroft},
  {Aussel}, {Bardelli}, {Bolzonella}, {Cappi}, {Caputi}, {Contini},
  {Finoguenov}, {Fruscione}, {Garilli}, {Impey}, {Iovino}, {Iwasawa},
  {Kampczyk}, {Kartaltepe}, {Kneib}, {Knobel}, {Kovac}, {Lamareille},
  {Leborgne}, {Le Brun}, {Le Fevre}, {Lilly}, {Maier}, {McCracken}, {Pello},
  {Peng}, {Perez-Montero}, {de Ravel}, {Sanders}, {Scodeggio}, {Scoville},
  {Tanaka}, {Taniguchi}, {Tasca}, {de la Torre}, {Tresse}, {Vergani}, \&
  {Zucca}}]{2010ApJ...716..348B}
{Brusa}, M., {et~al.} 2010, \apj, 716, 348

\bibitem[{{Bruzual} \& {Charlot}(2003)}]{2003MNRAS.344.1000B}
{Bruzual}, G., \& {Charlot}, S. 2003, \mnras, 344, 1000

\bibitem[{{Capak} {et~al.}(2007){Capak}, {Aussel}, {Ajiki}, {McCracken},
  {Mobasher}, {Scoville}, {Shopbell}, {Taniguchi}, {Thompson}, {Tribiano},
  {Sasaki}, {Blain}, {Brusa}, {Carilli}, {Comastri}, {Carollo}, {Cassata},
  {Colbert}, {Ellis}, {Elvis}, {Giavalisco}, {Green}, {Guzzo}, {Hasinger},
  {Ilbert}, {Impey}, {Jahnke}, {Kartaltepe}, {Kneib}, {Koda}, {Koekemoer},
  {Komiyama}, {Leauthaud}, {Le Fevre}, {Lilly}, {Liu}, {Massey}, {Miyazaki},
  {Murayama}, {Nagao}, {Peacock}, {Pickles}, {Porciani}, {Renzini}, {Rhodes},
  {Rich}, {Salvato}, {Sanders}, {Scarlata}, {Schiminovich}, {Schinnerer},
  {Scodeggio}, {Sheth}, {Shioya}, {Tasca}, {Taylor}, {Yan}, \&
  {Zamorani}}]{2007ApJS..172...99C}
{Capak}, P., {et~al.} 2007, \apjs, 172, 99

\bibitem[{{Casey} {et~al.}(2009){Casey}, {Chapman}, {Beswick}, {Biggs},
  {Blain}, {Hainline}, {Ivison}, {Muxlow}, \& {Smail}}]{2009MNRAS.399..121C}
{Casey}, C.~M., {et~al.} 2009, \mnras, 399, 121

\bibitem[{{Casey} {et~al.}(2012){Casey}, {Berta}, {B{\'e}thermin}, {Bock},
  {Bridge}, {Budynkiewicz}, {Burgarella}, {Chapin}, {Chapman}, {Clements},
  {Conley}, {Conselice}, {Cooray}, {Farrah}, {Hatziminaoglou}, {Ivison}, {le
  Floc'h}, {Lutz}, {Magdis}, {Magnelli}, {Oliver}, {Page}, {Pozzi},
  {Rigopoulou}, {Riguccini}, {Roseboom}, {Sanders}, {Scott}, {Seymour},
  {Valtchanov}, {Vieira}, {Viero}, \& {Wardlow}}]{2012ApJ...761..140C}
---. 2012, \apj, 761, 140

\bibitem[{{Chabrier}(2003)}]{2003PASP..115..763C}
{Chabrier}, G. 2003, \pasp, 115, 763

\bibitem[{{Chapman} {et~al.}(2004)}]{2004ApJ...614..671C}
{Chapman}, S.~C., {et~al.} 2004, \apj, 614, 671

\bibitem[{{Chapman} {et~al.}(2005)}]{2005ApJ...622..772C}
---. 2005, \apj, 622, 772

\bibitem[{{Chary} \& {Elbaz}(2001)}]{Chary:2001p1425}
{Chary}, R., \& {Elbaz}, D. 2001, \apj, 556, 562

\bibitem[{{Civano} {et~al.}(2011){Civano}, {Brusa}, {Comastri}, {Elvis},
  {Salvato}, {Zamorani}, {Capak}, {Fiore}, {Gilli}, {Hao}, {Ikeda}, {Kakazu},
  {Kartaltepe}, {Masters}, {Miyaji}, {Mignoli}, {Puccetti}, {Shankar},
  {Silverman}, {Vignali}, {Zezas}, \& {Koekemoer}}]{2011ApJ...741...91C}
{Civano}, F., {et~al.} 2011, \apj, 741, 91

\bibitem[{{Condon}(1992)}]{1992ARA&A..30..575C}
{Condon}, J.~J. 1992, \araa, 30, 575

\bibitem[{{Daddi} {et~al.}(2007){Daddi}, {Dickinson}, {Morrison}, {Chary},
  {Cimatti}, {Elbaz}, {Frayer}, {Renzini}, {Pope}, {Alexander}, {Bauer},
  {Giavalisco}, {Huynh}, {Kurk}, \& {Mignoli}}]{2007ApJ...670..156D}
{Daddi}, E., {et~al.} 2007, \apj, 670, 156

\bibitem[{{Dale} \& {Helou}(2002)}]{2002ApJ...576..159D}
{Dale}, D.~A., \& {Helou}, G. 2002, \apj, 576, 159

\bibitem[{{Diolaiti} {et~al.}(2000)}]{2000A&AS..147..335D}
{Diolaiti}, E., {et~al.} 2000, \aaps, 147, 335

\bibitem[{{Draine} \& {Li}(2007)}]{2007ApJ...657..810D}
{Draine}, B.~T., \& {Li}, A. 2007, \apj, 657, 810

\bibitem[{{Elbaz} {et~al.}(2007){Elbaz}, {Daddi}, {Le Borgne}, {Dickinson},
  {Alexander}, {Chary}, {Starck}, {Brandt}, {Kitzbichler}, {MacDonald},
  {Nonino}, {Popesso}, {Stern}, \& {Vanzella}}]{2007A&A...468...33E}
{Elbaz}, D., {et~al.} 2007, \aap, 468, 33

\bibitem[{{Elbaz} {et~al.}(2011){Elbaz}, {Dickinson}, {Hwang},
  {D{\'{\i}}az-Santos}, {Magdis}, {Magnelli}, {Le Borgne}, {Galliano},
  {Pannella}, {Chanial}, {Armus}, {Charmandaris}, {Daddi}, {Aussel}, {Popesso},
  {Kartaltepe}, {Altieri}, {Valtchanov}, {Coia}, {Dannerbauer}, {Dasyra},
  {Leiton}, {Mazzarella}, {Alexander}, {Buat}, {Burgarella}, {Chary}, {Gilli},
  {Ivison}, {Juneau}, {Le Floc'h}, {Lutz}, {Morrison}, {Mullaney}, {Murphy},
  {Pope}, {Scott}, {Brodwin}, {Calzetti}, {Cesarsky}, {Charlot}, {Dole},
  {Eisenhardt}, {Ferguson}, {F{\"o}rster Schreiber}, {Frayer}, {Giavalisco},
  {Huynh}, {Koekemoer}, {Papovich}, {Reddy}, {Surace}, {Teplitz}, {Yun}, \&
  {Wilson}}]{2011A&A...533A.119E}
---. 2011, A\&A, 533, A119

\bibitem[{{Griffin} {et~al.}(2010)}]{2010A&A...518L...3G}
{Griffin}, M.~J., {et~al.} 2010, \aap, 518, L3

\bibitem[{{Helou} {et~al.}(1985){Helou}, {Soifer}, \&
  {Rowan-Robinson}}]{1985ApJ...298L...7H}
{Helou}, G., {Soifer}, B.~T., \& {Rowan-Robinson}, M. 1985, \apjl, 298, L7

\bibitem[{{Hung} {et~al.}(2013){Hung}, {Sanders}, {Casey}, {Lee}, {Barnes},
  {Capak}, {Kartaltepe}, {Koss}, {Larson}, {Le Floc'h}, {Lockhart}, {Man},
  {Mann}, {Riguccini}, {Scoville}, \& {Symeonidis}}]{2013arXiv1309.4459H}
{Hung}, C.-L., {et~al.} 2013, ArXiv e-prints

\bibitem[{{Ilbert} {et~al.}(2009){Ilbert}, {Capak}, {Salvato}, {Aussel},
  {McCracken}, {Sanders}, {Scoville}, {Kartaltepe}, {Arnouts}, {Le Floc'h},
  {Mobasher}, {Taniguchi}, {Lamareille}, {Leauthaud}, {Sasaki}, {Thompson},
  {Zamojski}, {Zamorani}, {Bardelli}, {Bolzonella}, {Bongiorno}, {Brusa},
  {Caputi}, {Carollo}, {Contini}, {Cook}, {Coppa}, {Cucciati}, {de la Torre},
  {de Ravel}, {Franzetti}, {Garilli}, {Hasinger}, {Iovino}, {Kampczyk},
  {Kneib}, {Knobel}, {Kovac}, {Le Borgne}, {Le Brun}, {F{\`e}vre}, {Lilly},
  {Looper}, {Maier}, {Mainieri}, {Mellier}, {Mignoli}, {Murayama}, {Pell{\`o}},
  {Peng}, {P{\'e}rez-Montero}, {Renzini}, {Ricciardelli}, {Schiminovich},
  {Scodeggio}, {Shioya}, {Silverman}, {Surace}, {Tanaka}, {Tasca}, {Tresse},
  {Vergani}, \& {Zucca}}]{2009ApJ...690.1236I}
{Ilbert}, O., {et~al.} 2009, ApJ, 690, 1236

\bibitem[{{Ilbert} {et~al.}(2010){Ilbert}, {Salvato}, {Le Floc'h}, {Aussel},
  {Capak}, {McCracken}, {Mobasher}, {Kartaltepe}, {Scoville}, {Sanders},
  {Arnouts}, {Bundy}, {Cassata}, {Kneib}, {Koekemoer}, {Le F{\`e}vre}, {Lilly},
  {Surace}, {Taniguchi}, {Tasca}, {Thompson}, {Tresse}, {Zamojski}, {Zamorani},
  \& {Zucca}}]{2010ApJ...709..644I}
---. 2010, \apj, 709, 644

\bibitem[{{Kennicutt}(1998)}]{1998ARA&A..36..189K}
{Kennicutt}, Jr., R.~C. 1998, \araa, 36, 189

\bibitem[{{Le Floc'h} {et~al.}(2005){Le Floc'h}, {Papovich}, {Dole}, {Bell},
  {Lagache}, {Rieke}, {Egami}, {P{\'e}rez-Gonz{\'a}lez}, {Alonso-Herrero},
  {Rieke}, {Blaylock}, {Engelbracht}, {Gordon}, {Hines}, {Misselt}, {Morrison},
  \& {Mould}}]{2005ApJ...632..169L}
{Le Floc'h}, E., {et~al.} 2005, \apj, 632, 169

\bibitem[{{Le Floc'h} {et~al.}(2009){Le Floc'h}, {Aussel}, {Ilbert},
  {Riguccini}, {Frayer}, {Salvato}, {Arnouts}, {Surace}, {Feruglio},
  {Rodighiero}, {Capak}, {Kartaltepe}, {Heinis}, {Sheth}, {Yan}, {McCracken},
  {Thompson}, {Sanders}, {Scoville}, \& {Koekemoer}}]{2009ApJ...703..222L}
---. 2009, \apj, 703, 222

\bibitem[{{Lee} {et~al.}(2010){Lee}, {Le Floc'h}, {Sanders}, {Frayer},
  {Arnouts}, {Ilbert}, {Aussel}, {Salvato}, {Scoville}, \&
  {Kartaltepe}}]{2010ApJ...717..175L}
{Lee}, N., {et~al.} 2010, \apj, 717, 175

\bibitem[{{Lilly} {et~al.}(2007){Lilly}, {Le F{\`e}vre}, {Renzini}, {Zamorani},
  {Scodeggio}, {Contini}, {Carollo}, {Hasinger}, {Kneib}, {Iovino}, {Le Brun},
  {Maier}, {Mainieri}, {Mignoli}, {Silverman}, {Tasca}, {Bolzonella},
  {Bongiorno}, {Bottini}, {Capak}, {Caputi}, {Cimatti}, {Cucciati}, {Daddi},
  {Feldmann}, {Franzetti}, {Garilli}, {Guzzo}, {Ilbert}, {Kampczyk}, {Kovac},
  {Lamareille}, {Leauthaud}, {Borgne}, {McCracken}, {Marinoni}, {Pello},
  {Ricciardelli}, {Scarlata}, {Vergani}, {Sanders}, {Schinnerer}, {Scoville},
  {Taniguchi}, {Arnouts}, {Aussel}, {Bardelli}, {Brusa}, {Cappi}, {Ciliegi},
  {Finoguenov}, {Foucaud}, {Franceschini}, {Halliday}, {Impey}, {Knobel},
  {Koekemoer}, {Kurk}, {Maccagni}, {Maddox}, {Marano}, {Marconi}, {Meneux},
  {Mobasher}, {Moreau}, {Peacock}, {Porciani}, {Pozzetti}, {Scaramella},
  {Schiminovich}, {Shopbell}, {Smail}, {Thompson}, {Tresse}, {Vettolani},
  {Zanichelli}, \& {Zucca}}]{2007ApJS..172...70L}
{Lilly}, S.~J., {et~al.} 2007, \apjs, 172, 70

\bibitem[{{Lutz} {et~al.}(2011){Lutz}, {Poglitsch}, {Altieri}, {Andreani},
  {Aussel}, {Berta}, {Bongiovanni}, {Brisbin}, {Cava}, {Cepa}, {Cimatti},
  {Daddi}, {Dominguez-Sanchez}, {Elbaz}, {F{\"o}rster Schreiber}, {Genzel},
  {Grazian}, {Gruppioni}, {Harwit}, {Le Floc'h}, {Magdis}, {Magnelli},
  {Maiolino}, {Nordon}, {P{\'e}rez Garc{\'{\i}}a}, {Popesso}, {Pozzi},
  {Riguccini}, {Rodighiero}, {Saintonge}, {Sanchez Portal}, {Santini}, {Shao},
  {Sturm}, {Tacconi}, {Valtchanov}, {Wetzstein}, \&
  {Wieprecht}}]{2011A&A...532A..90L}
{Lutz}, D., {et~al.} 2011, \aap, 532, A90

\bibitem[{{Magdis} {et~al.}(2011){Magdis}, {Elbaz}, {Dickinson}, {Hwang},
  {Charmandaris}, {Armus}, {Daddi}, {Le Floc'h}, {Aussel}, {Dannerbauer},
  {Rigopoulou}, {Buat}, {Morrison}, {Mullaney}, {Lutz}, {Scott}, {Coia},
  {Pope}, {Pannella}, {Altieri}, {Burgarella}, {Bethermin}, {Dasyra},
  {Kartaltepe}, {Leiton}, {Magnelli}, {Popesso}, \&
  {Valtchanov}}]{2011A&A...534A..15M}
{Magdis}, G.~E., {et~al.} 2011, A\&A, 534, A15

\bibitem[{{Magnelli} {et~al.}(2009)}]{2009A&A...496...57M}
{Magnelli}, B., {et~al.} 2009, \aap, 496, 57

\bibitem[{{Malkan} \& {Sargent}(1982)}]{1982ApJ...254...22M}
{Malkan}, M.~A., \& {Sargent}, W.~L.~W. 1982, \apj, 254, 22

\bibitem[{{McCracken} {et~al.}(2010){McCracken}, {Capak}, {Salvato}, {Aussel},
  {Thompson}, {Daddi}, {Sanders}, {Kneib}, {Willott}, {Mancini}, {Renzini},
  {Cook}, {Le F{\`e}vre}, {Ilbert}, {Kartaltepe}, {Koekemoer}, {Mellier},
  {Murayama}, {Scoville}, {Shioya}, \& {Tanaguchi}}]{McCracken:2009p2464}
{McCracken}, H.~J., {et~al.} 2010, \apj, 708, 202

\bibitem[{{McCracken} {et~al.}(2012){McCracken}, {Milvang-Jensen}, {Dunlop},
  {Franx}, {Fynbo}, {Le F{\`e}vre}, {Holt}, {Caputi}, {Goranova}, {Buitrago},
  {Emerson}, {Freudling}, {Hudelot}, {L{\'o}pez-Sanjuan}, {Magnard}, {Mellier},
  {M{\o}ller}, {Nilsson}, {Sutherland}, {Tasca}, \&
  {Zabl}}]{2012A&A...544A.156M}
---. 2012, \aap, 544, A156

\bibitem[{{Micha{\l}owski} {et~al.}(2012)}]{2012A&A...541A..85M}
{Micha{\l}owski}, M.~J., {et~al.} 2012, \aap, 541, A85

\bibitem[{{Muzzin} {et~al.}(2013){Muzzin}, {Marchesini}, {Stefanon}, {Franx},
  {Milvang-Jensen}, {Dunlop}, {Fynbo}, {Brammer}, {Labb{\'e}}, \& {van
  Dokkum}}]{2013ApJS..206....8M}
{Muzzin}, A., {et~al.} 2013, \apjs, 206, 8

\bibitem[{{Noeske} {et~al.}(2007){Noeske}, {Weiner}, {Faber}, {Papovich},
  {Koo}, {Somerville}, {Bundy}, {Conselice}, {Newman}, {Schiminovich}, {Le
  Floc'h}, {Coil}, {Rieke}, {Lotz}, {Primack}, {Barmby}, {Cooper}, {Davis},
  {Ellis}, {Fazio}, {Guhathakurta}, {Huang}, {Kassin}, {Martin}, {Phillips},
  {Rich}, {Small}, {Willmer}, \& {Wilson}}]{2007ApJ...660L..43N}
{Noeske}, K.~G., {et~al.} 2007, ApJ, 660, L43

\bibitem[{{Nordon} {et~al.}(2012){Nordon}, {Lutz}, {Genzel}, {Berta}, {Wuyts},
  {Magnelli}, {Altieri}, {Andreani}, {Aussel}, {Bongiovanni}, {Cepa},
  {Cimatti}, {Daddi}, {Fadda}, {F{\"o}rster Schreiber}, {Lagache}, {Maiolino},
  {P{\'e}rez Garc{\'{\i}}a}, {Poglitsch}, {Popesso}, {Pozzi}, {Rodighiero},
  {Rosario}, {Saintonge}, {Sanchez-Portal}, {Santini}, {Sturm}, {Tacconi},
  {Valtchanov}, \& {Yan}}]{2012ApJ...745..182N}
{Nordon}, R., {et~al.} 2012, \apj, 745, 182

\bibitem[{{Oliver} {et~al.}(2012){Oliver}, {Bock}, {Altieri}, {Amblard},
  {Arumugam}, {Aussel}, {Babbedge}, {Beelen}, {B{\'e}thermin}, {Blain},
  {Boselli}, {Bridge}, {Brisbin}, {Buat}, {Burgarella},
  {Castro-Rodr{\'{\i}}guez}, {Cava}, {Chanial}, {Cirasuolo}, {Clements},
  {Conley}, {Conversi}, {Cooray}, {Dowell}, {Dubois}, {Dwek}, {Dye}, {Eales},
  {Elbaz}, {Farrah}, {Feltre}, {Ferrero}, {Fiolet}, {Fox}, {Franceschini},
  {Gear}, {Giovannoli}, {Glenn}, {Gong}, {Gonz{\'a}lez Solares}, {Griffin},
  {Halpern}, {Harwit}, {Hatziminaoglou}, {Heinis}, {Hurley}, {Hwang}, {Hyde},
  {Ibar}, {Ilbert}, {Isaak}, {Ivison}, {Lagache}, {Le Floc'h}, {Levenson},
  {Faro}, {Lu}, {Madden}, {Maffei}, {Magdis}, {Mainetti}, {Marchetti},
  {Marsden}, {Marshall}, {Mortier}, {Nguyen}, {O'Halloran}, {Omont}, {Page},
  {Panuzzo}, {Papageorgiou}, {Patel}, {Pearson}, {P{\'e}rez-Fournon}, {Pohlen},
  {Rawlings}, {Raymond}, {Rigopoulou}, {Riguccini}, {Rizzo}, {Rodighiero},
  {Roseboom}, {Rowan-Robinson}, {S{\'a}nchez Portal}, {Schulz}, {Scott},
  {Seymour}, {Shupe}, {Smith}, {Stevens}, {Symeonidis}, {Trichas}, {Tugwell},
  {Vaccari}, {Valtchanov}, {Vieira}, {Viero}, {Vigroux}, {Wang}, {Ward},
  {Wardlow}, {Wright}, {Xu}, \& {Zemcov}}]{2012MNRAS.424.1614O}
{Oliver}, S.~J., {et~al.} 2012, \mnras, 424, 1614

\bibitem[{{Papovich} {et~al.}(2007){Papovich}, {Rudnick}, {Le Floc'h}, {van
  Dokkum}, {Rieke}, {Taylor}, {Armus}, {Gawiser}, {Huang}, {Marcillac}, \&
  {Franx}}]{Papovich:2007p38}
{Papovich}, C., {et~al.} 2007, \apj, 668, 45

\bibitem[{{Pilbratt} {et~al.}(2010){Pilbratt}, {Riedinger}, {Passvogel},
  {Crone}, {Doyle}, {Gageur}, {Heras}, {Jewell}, {Metcalfe}, {Ott}, \&
  {Schmidt}}]{2010A&A...518L...1P}
{Pilbratt}, G.~L., {et~al.} 2010, \aap, 518, L1

\bibitem[{{Poglitsch} {et~al.}(2010){Poglitsch}, {Waelkens}, {Geis},
  {Feuchtgruber}, {Vandenbussche}, {Rodriguez}, {Krause}, {Renotte}, {van
  Hoof}, {Saraceno}, {Cepa}, {Kerschbaum}, {Agn{\`e}se}, {Ali}, {Altieri},
  {Andreani}, {Augueres}, {Balog}, {Barl}, {Bauer}, {Belbachir}, {Benedettini},
  {Billot}, {Boulade}, {Bischof}, {Blommaert}, {Callut}, {Cara}, {Cerulli},
  {Cesarsky}, {Contursi}, {Creten}, {De Meester}, {Doublier}, {Doumayrou},
  {Duband}, {Exter}, {Genzel}, {Gillis}, {Gr{\"o}zinger}, {Henning},
  {Herreros}, {Huygen}, {Inguscio}, {Jakob}, {Jamar}, {Jean}, {de Jong},
  {Katterloher}, {Kiss}, {Klaas}, {Lemke}, {Lutz}, {Madden}, {Marquet},
  {Martignac}, {Mazy}, {Merken}, {Montfort}, {Morbidelli}, {M{\"u}ller},
  {Nielbock}, {Okumura}, {Orfei}, {Ottensamer}, {Pezzuto}, {Popesso},
  {Putzeys}, {Regibo}, {Reveret}, {Royer}, {Sauvage}, {Schreiber}, {Stegmaier},
  {Schmitt}, {Schubert}, {Sturm}, {Thiel}, {Tofani}, {Vavrek}, {Wetzstein},
  {Wieprecht}, \& {Wiezorrek}}]{2010A&A...518L...2P}
{Poglitsch}, A., {et~al.} 2010, \aap, 518, L2

\bibitem[{{Polletta} {et~al.}(2007){Polletta}, {Tajer}, {Maraschi},
  {Trinchieri}, {Lonsdale}, {Chiappetti}, {Andreon}, {Pierre}, {Le F{\`e}vre},
  {Zamorani}, {Maccagni}, {Garcet}, {Surdej}, {Franceschini}, {Alloin},
  {Shupe}, {Surace}, {Fang}, {Rowan-Robinson}, {Smith}, \&
  {Tresse}}]{2007ApJ...663...81P}
{Polletta}, M., {et~al.} 2007, \apj, 663, 81

\bibitem[{{Pope} {et~al.}(2008){Pope}, {Chary}, {Alexander}, {Armus},
  {Dickinson}, {Elbaz}, {Frayer}, {Scott}, \& {Teplitz}}]{2008ApJ...675.1171P}
{Pope}, A., {et~al.} 2008, \apj, 675, 1171

\bibitem[{{Rieke} {et~al.}(2009)}]{2009ApJ...692..556R}
{Rieke}, G.~H., {et~al.} 2009, \apj, 692, 556

\bibitem[{{Rodighiero} {et~al.}(2011){Rodighiero}, {Daddi}, {Baronchelli},
  {Cimatti}, {Renzini}, {Aussel}, {Popesso}, {Lutz}, {Andreani}, {Berta},
  {Cava}, {Elbaz}, {Feltre}, {Fontana}, {F{\"o}rster Schreiber},
  {Franceschini}, {Genzel}, {Grazian}, {Gruppioni}, {Ilbert}, {Le Floch},
  {Magdis}, {Magliocchetti}, {Magnelli}, {Maiolino}, {McCracken}, {Nordon},
  {Poglitsch}, {Santini}, {Pozzi}, {Riguccini}, {Tacconi}, {Wuyts}, \&
  {Zamorani}}]{2011ApJ...739L..40R}
{Rodighiero}, G., {et~al.} 2011, ApJ, 739, L40

\bibitem[{{Roseboom} {et~al.}(2010){Roseboom}, {Oliver}, {Kunz}, {Altieri},
  {Amblard}, {Arumugam}, {Auld}, {Aussel}, {Babbedge}, {B{\'e}thermin},
  {Blain}, {Bock}, {Boselli}, {Brisbin}, {Buat}, {Burgarella},
  {Castro-Rodr{\'{\i}}guez}, {Cava}, {Chanial}, {Chapin}, {Clements}, {Conley},
  {Conversi}, {Cooray}, {Dowell}, {Dwek}, {Dye}, {Eales}, {Elbaz}, {Farrah},
  {Fox}, {Franceschini}, {Gear}, {Glenn}, {Solares}, {Griffin}, {Halpern},
  {Harwit}, {Hatziminaoglou}, {Huang}, {Ibar}, {Isaak}, {Ivison}, {Lagache},
  {Levenson}, {Lu}, {Madden}, {Maffei}, {Mainetti}, {Marchetti}, {Marsden},
  {Mortier}, {Nguyen}, {O'Halloran}, {Omont}, {Page}, {Panuzzo},
  {Papageorgiou}, {Patel}, {Pearson}, {P{\'e}rez-Fournon}, {Pohlen},
  {Rawlings}, {Raymond}, {Rigopoulou}, {Rizzo}, {Rowan-Robinson}, {Portal},
  {Schulz}, {Scott}, {Seymour}, {Shupe}, {Smith}, {Stevens}, {Symeonidis},
  {Trichas}, {Tugwell}, {Vaccari}, {Valtchanov}, {Vieira}, {Vigroux}, {Wang},
  {Ward}, {Wright}, {Xu}, \& {Zemcov}}]{2010MNRAS.409...48R}
{Roseboom}, I.~G., {et~al.} 2010, \mnras, 409, 48

\bibitem[{{Roseboom} {et~al.}(2012){Roseboom}, {Ivison}, {Greve}, {Amblard},
  {Arumugam}, {Auld}, {Aussel}, {Bethermin}, {Blain}, {Bock}, {Boselli},
  {Brisbin}, {Buat}, {Burgarella}, {Castro-Rodr{\'{\i}}guez}, {Cava},
  {Chanial}, {Chapin}, {Chapman}, {Clements}, {Conley}, {Conversi}, {Cooray},
  {Dowell}, {Dunlop}, {Dwek}, {Eales}, {Elbaz}, {Farrah}, {Franceschini},
  {Glenn}, {Griffin}, {Halpern}, {Hatziminaoglou}, {Ibar}, {Isaak}, {Lagache},
  {Levenson}, {Lu}, {Madden}, {Maffei}, {Mainetti}, {Marchetti}, {Marsden},
  {Morrison}, {Mortier}, {Nguyen}, {O'Halloran}, {Oliver}, {Omont}, {Page},
  {Panuzzo}, {Papageorgiou}, {Pearson}, {P{\'e}rez-Fournon}, {Pohlen},
  {Rawlings}, {Raymond}, {Rigopoulou}, {Rizzo}, {Rodighiero}, {Rowan-Robinson},
  {Schulz}, {Scott}, {Seymour}, {Shupe}, {Smith}, {Stevens}, {Symeonidis},
  {Trichas}, {Tugwell}, {Vaccari}, {Valtchanov}, {Vieira}, {Viero}, {Vigroux},
  {Wardlow}, {Wang}, {Wright}, {Xu}, \& {Zemcov}}]{2012MNRAS.419.2758R}
---. 2012, \mnras, 419, 2758

\bibitem[{{Salpeter}(1955)}]{1955ApJ...121..161S}
{Salpeter}, E.~E. 1955, \apj, 121, 161

\bibitem[{{Salvato} {et~al.}(2011){Salvato}, {Ilbert}, {Hasinger}, {Rau},
  {Civano}, {Zamorani}, {Brusa}, {Elvis}, {Vignali}, {Aussel}, {Comastri},
  {Fiore}, {Le Floc'h}, {Mainieri}, {Bardelli}, {Bolzonella}, {Bongiorno},
  {Capak}, {Caputi}, {Cappelluti}, {Carollo}, {Contini}, {Garilli}, {Iovino},
  {Fotopoulou}, {Fruscione}, {Gilli}, {Halliday}, {Kneib}, {Kakazu},
  {Kartaltepe}, {Koekemoer}, {Kovac}, {Ideue}, {Ikeda}, {Impey}, {Le Fevre},
  {Lamareille}, {Lanzuisi}, {Le Borgne}, {Le Brun}, {Lilly}, {Maier},
  {Manohar}, {Masters}, {McCracken}, {Messias}, {Mignoli}, {Mobasher}, {Nagao},
  {Pello}, {Puccetti}, {Perez-Montero}, {Renzini}, {Sargent}, {Sanders},
  {Scodeggio}, {Scoville}, {Shopbell}, {Silvermann}, {Taniguchi}, {Tasca},
  {Tresse}, {Trump}, \& {Zucca}}]{2011ApJ...742...61S}
{Salvato}, M., {et~al.} 2011, \apj, 742, 61

\bibitem[{{Sanders} \& {Mirabel}(1996)}]{1996ARA&A..34..749S}
{Sanders}, D.~B., \& {Mirabel}, I.~F. 1996, \araa, 34, 749

\bibitem[{{Sanders} {et~al.}(2007){Sanders}, {Salvato}, {Aussel}, {Ilbert},
  {Scoville}, {Surace}, {Frayer}, {Sheth}, {Helou}, {Brooke}, {Bhattacharya},
  {Yan}, {Kartaltepe}, {Barnes}, {Blain}, {Calzetti}, {Capak}, {Carilli},
  {Carollo}, {Comastri}, {Daddi}, {Ellis}, {Elvis}, {Fall}, {Franceschini},
  {Giavalisco}, {Hasinger}, {Impey}, {Koekemoer}, {Le F{\`e}vre}, {Lilly},
  {Liu}, {McCracken}, {Mobasher}, {Renzini}, {Rich}, {Schinnerer}, {Shopbell},
  {Taniguchi}, {Thompson}, {Urry}, \& {Williams}}]{2007ApJS..172...86S}
{Sanders}, D.~B., {et~al.} 2007, \apjs, 172, 86

\bibitem[{{Schinnerer} {et~al.}(2010){Schinnerer}, {Sargent}, {Bondi}, {Smol{\v
  c}i{\'c}}, {Datta}, {Carilli}, {Bertoldi}, {Blain}, {Ciliegi}, {Koekemoer},
  \& {Scoville}}]{2010ApJS..188..384S}
{Schinnerer}, E., {et~al.} 2010, \apjs, 188, 384

\bibitem[{{Scott} {et~al.}(2008){Scott}, {Austermann}, {Perera}, {Wilson},
  {Aretxaga}, {Bock}, {Hughes}, {Kang}, {Kim}, {Mauskopf}, {Sanders},
  {Scoville}, \& {Yun}}]{2008MNRAS.385.2225S}
{Scott}, K.~S., {et~al.} 2008, \mnras, 385, 2225

\bibitem[{{Scoville} {et~al.}(2007){Scoville}, {Aussel}, {Brusa}, {Capak},
  {Carollo}, {Elvis}, {Giavalisco}, {Guzzo}, {Hasinger}, {Impey}, {Kneib},
  {LeFevre}, {Lilly}, {Mobasher}, {Renzini}, {Rich}, {Sanders}, {Schinnerer},
  {Schminovich}, {Shopbell}, {Taniguchi}, \& {Tyson}}]{2007ApJS..172....1S}
{Scoville}, N., {et~al.} 2007, \apjs, 172, 1

\bibitem[{{Shields}(1978)}]{1978Natur.272..706S}
{Shields}, G.~A. 1978, \nat, 272, 706

\bibitem[{{Siebenmorgen} \& {Kr{\"u}gel}(2007)}]{Siebenmorgen:2007p1876}
{Siebenmorgen}, R., \& {Kr{\"u}gel}, E. 2007, \aap, 461, 445

\bibitem[{{Symeonidis} {et~al.}(2013){Symeonidis}, {Vaccari}, {Berta}, {Page},
  {Lutz}, {Arumugam}, {Aussel}, {Bock}, {Boselli}, {Buat}, {Capak}, {Clements},
  {Conley}, {Conversi}, {Cooray}, {Dowell}, {Farrah}, {Franceschini},
  {Giovannoli}, {Glenn}, {Griffin}, {Hatziminaoglou}, {Hwang}, {Ibar},
  {Ilbert}, {Ivison}, {Floc'h}, {Lilly}, {Kartaltepe}, {Magnelli}, {Magdis},
  {Marchetti}, {Nguyen}, {Nordon}, {O'Halloran}, {Oliver}, {Omont},
  {Papageorgiou}, {Patel}, {Pearson}, {P{\'e}rez-Fournon}, {Pohlen}, {Popesso},
  {Pozzi}, {Rigopoulou}, {Riguccini}, {Rosario}, {Roseboom}, {Rowan-Robinson},
  {Salvato}, {Schulz}, {Scott}, {Seymour}, {Shupe}, {Smith}, {Valtchanov},
  {Wang}, {Xu}, {Zemcov}, \& {Wuyts}}]{2013MNRAS.431.2317S}
{Symeonidis}, M., {et~al.} 2013, \mnras, 431, 2317

\bibitem[{{Taniguchi} {et~al.}(2007){Taniguchi}, {Scoville}, {Murayama},
  {Sanders}, {Mobasher}, {Aussel}, {Capak}, {Ajiki}, {Miyazaki}, {Komiyama},
  {Shioya}, {Nagao}, {Sasaki}, {Koda}, {Carilli}, {Giavalisco}, {Guzzo},
  {Hasinger}, {Impey}, {LeFevre}, {Lilly}, {Renzini}, {Rich}, {Schinnerer},
  {Shopbell}, {Kaifu}, {Karoji}, {Arimoto}, {Okamura}, \&
  {Ohta}}]{2007ApJS..172....9T}
{Taniguchi}, Y., {et~al.} 2007, \apjs, 172, 9

\bibitem[{{Tielens} {et~al.}(1999){Tielens}, {Hony}, {van Kerckhoven}, \&
  {Peeters}}]{1999ESASP.427..579T}
{Tielens}, A.~G.~G.~M., {Hony}, S., {van Kerckhoven}, C., \& {Peeters}, E.
  1999, in ESA Special Publication, Vol. 427, The Universe as Seen by ISO, ed.
  P.~{Cox} \& M.~{Kessler}, 579

\bibitem[{{Treister} {et~al.}(2009){Treister}, {Cardamone}, {Schawinski},
  {Urry}, {Gawiser}, {Virani}, {Lira}, {Kartaltepe}, {Damen}, {Taylor}, {Le
  Floc'h}, {Justham}, \& {Koekemoer}}]{2009ApJ...706..535T}
{Treister}, E., {et~al.} 2009, \apj, 706, 535

\bibitem[{{Treister} {et~al.}(2010)}]{2010ApJ...722L.238T}
---. 2010, \apjl, 722, L238

\end{thebibliography}

\end{document}